\documentclass[aps,prb,twocolumn,longbibliography,floatfix]{revtex4-2}
\usepackage{float,amsmath,amssymb,bbm,mathrsfs,bm,braket,color,graphicx,comment,amsfonts,dsfont,mathrsfs}
\usepackage{dcolumn}% Align table columns on decimal point
\usepackage{bm}% bold math
\usepackage{amsmath}
\usepackage{amssymb}
\usepackage[usenames,dvipsnames]{xcolor} % this is used for the comment colors
\usepackage{mathtools}
\usepackage[colorlinks=true,citecolor=blue,linkcolor=blue]{hyperref}
\usepackage[caption=false]{subfig}
\usepackage{hyperref}
\usepackage{upgreek}
\usepackage{esint}
\usepackage{soul}
\usepackage[normalem]{ ulem } 
\usepackage{slashed}
\usepackage{float}

\newcommand{\bea}{\begin{eqnarray}}
\newcommand{\eea}{\end{eqnarray}}

\newcommand{\no}{\nonumber}

\begin{document}

\title{Fully Tunable Fano Resonances in Chiral Electronic Transport}

\author {Ai-Ying Ye}
\author{Zhao Yang Zeng}\email{zyzeng@hunnu.edu.cn}
\address {Department of Physics, Jiangxi Normal University, Nanchang 330022, China} 

\begin{abstract}
Fano resonance is believed to arise when a direct path interferes with a resonant path. We demonstrate that this is not true for chiral electronic transmission without additional direct paths. To address the Fano effect in chiral electronic transport, we suggest an electronic Mach-Zehnder-Fano interferometer, which integrates a quantum dot into an electronic Mach-Zehnder interferometer. Due to the absence of backscattering in chiral electronic transport, Fano resonances can be fully adjusted by an external magnetic flux in the transmission, linear conductance, differential conductance and differential shot noise of chiral electrons. Even the current and shot noise for a symmetric interferometer with two arms of the same length exhibit fully controllable resonances and distinct Fano characteristics. In particular, all the profiles in the various transport spectra follow the same evolution pattern in an evolution cycle that is resistant to changes in the device's defining parameters.

\end{abstract}

\maketitle
\date{\today}

\section{Introduction}
\label{sec.intr}
%%%%%%%%%%%%%%%%%%%%%%%%%%%%%%%%%%%%%%%%%%%%%%%%%%

Fano resonance\cite{Fano,Miroshnichenko,Limonov}, characterized by asymmetric line profiles, is of fundamental and practical interest. It results from interference between one resonant path and one nonresonant (direct) path. The Fano profile is defined as $(\delta+q)^2/(\delta^2+1)$, where $\delta$ represents the dimensionless detuning and $q$ is the asymmetry parameter. Fano line profiles have been observed in the characteristic spectra of a number of systems, including photoabsorption in atoms\cite{Beutler}, neutrons\cite{Adair} and electrons\cite{Simpson} scattering, Raman scattering\cite{Cerdeira}, photoabsorption in quantum wells\cite{Feist}, scanning tunneling microscopy\cite{Madhavan}, and conductance through quantum dots (\textcolor{blue}{QD}s)\cite{Gores}.

Resonant tunneling through \textcolor{blue}{QD} devices has drawn a great deal of attention in the past decades. Conductance peaks in \textcolor{blue}{QD}s exhibit the standard Lorentzian profile of a Breit-Wigner resonance $1/(\delta^2+1)$ \cite{Meirav,Beenakker}, despite strong electron-electron interactions in \textcolor{blue}{QD}s. One may expect Fano-type line shapes in the transmission through a quantum wire side-coupled to a localized resonant mode\cite{Porod,Tekman,Shao,Nockel}. The first surprising discovery of the asymmetric line shape of differential conductance occurred in the study of Kondo resonance for tunneling into a surface magnetic atom \cite{Madhavan}. G\"{o}res et al.\cite{Gores} conducted a systematic experimental study on the Fano effect in electronic transport through a \textcolor{blue}{QD}, an electron droplet generated in a two-dimensional electron gas by electrostatic gating, which permits adjusting of resonant width. However, the transition of the conductance dip into a peak by altering the magnetic field is not well understood, and the origin of the nonresonant route is unclear\cite{Clerk,Gores}. This work sparked subsequent investigations on Fano resonances in the conductances through an Aharonov-Bohm(\textcolor{blue} {AB})ring  with a \textcolor{blue}{QD} embedded in one arm\cite{Kobayashi1,Kobayashi2,Aharony} and a quantum wire with a side-coupled \textcolor{blue} {QD}\cite{Kobayashi3,Johnson,Fuhrer} in the Coulomb blockade regime, as well as in the Kondo regime\cite{Bulka,Hofstetter,Torio,Aligia,Sato,Kang}.
 
 There has been little investigation into Fano resonances in chiral electronic transport thus far. Backscattering is entirely suppressed for chiral electrons, and as a consequence, it is meaningless to embed a \textcolor{blue}{QD} in a chiral quantum wire. Transmission probability through a chiral channel with a side-coupled resonant mode is always unity, irrespective of the electron's incident energy, and thus no Fano resonance can be expected. It is in striking contrast to the nonchiral case where the transmission probability develops a Fano dip at resonance \cite{Shao,Torio,Aligia,Kang} due to resonant reflection. However, if we introduce another chiral mode and allow electrons to tunnel into it, electrons will have two paths to choose from, and interference between these two channels could result in an unexpected Fano effect. To achieve this goal, we assemble the aforementioned elements into an electronic Mach-Zehnder-Fano interferometer (\textcolor{blue}{MZFI}), which may be the most straightforward configuration to examine the Fano effect  in chiral electron transport. The fact that each chiral electron only travels through the interferometer once simplifies the theoretical analysis and unravels the underlying physics more clearly than any nonchiral interferometer, where electrons traverse forth and back many times before departing.
 
 The rest of the paper is organized as follows. In Sec.~\ref{Mo&Form}, we describe the model of a \textcolor{blue}{MZFI} connected to four metallic contacts acting as electron sources and drains, formulate it in Hamiltonian language and present an explicit expression to calculate the current flowing into a drain contact. Sec.~\ref{Ana:Fano} is devoted to the analysis of Fano resonances in various transport 
 spectra, including transmission, linear conductance, differential conductance, current, shot noise and differential shot noise. Surprisingly, all the transport spectra of a symmetric \textcolor{blue}{MZFI} exhibit fully tunable Fano resonances and perform the same evolution pattern within a period as a function of the external magnetic flux. The effect of asymmetry in the lengths of two arms of the interferometer and interacting \textcolor{blue}{QD} in the Coulomb blockade and Kondo regime is analyzed and elucidated in Sec.~\ref{Dis}.
 We argue that, not being perfect, Fano resonances still persist and are fully tunable under these circumstances. Concluding remarks and possible further investigations in the chiral Fano effect are given in Sec.~\ref{Con}.     
 
 The calculations in this work were tedious and quilt involved. We have relegated many of the details to four appendixes. In Appendix~\ref{App:A} we present a detailed derivation of the current flowing into a drain contact in the absence of contact 
 Hamiltonian using the equation-of-motion method and the Keldysh equation. A standard Keldysh formalism to recover the current expression 
 in the presence of contact Hamiltonian is presented as a complementary in Appendix~\ref{App:B}. In Appendix~\ref{App:C} we justify the
 assumption $\theta(0)=1/2$  made in Appendix~\ref{App:A(3)} from the scattering matrix approach. Scattering matrix and shot noise of the device are derived based on the Fisher-Lee relation\cite{Fisher,Datta}  in Appendix~\ref{App:D}.  

\section{Model and Formulation}
\label{Mo&Form}

\begin{figure}
\includegraphics[width=0.8\columnwidth]{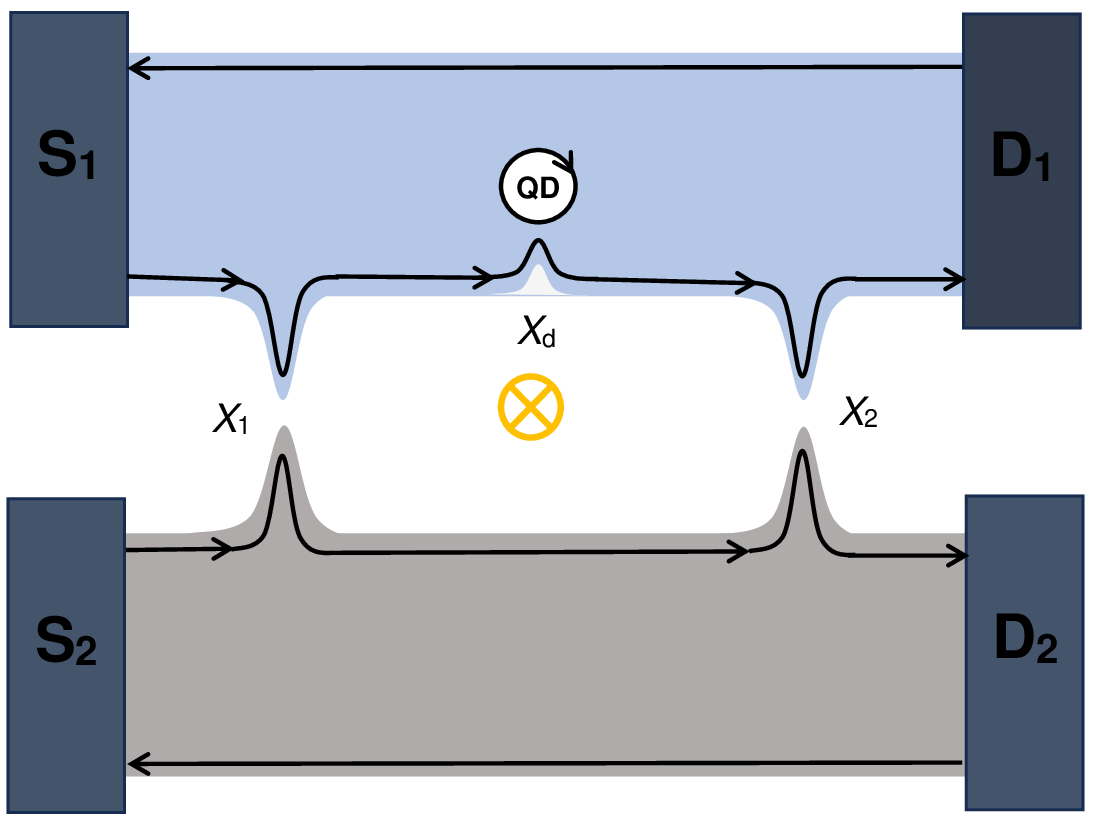}
\caption{\label{fig1}  Schematic illustration of a four-terminal electronic \textcolor{blue} {MZFI}: Two chiral edge states, represented by arrowhead lines, are formed at the boundaries of two quantum hall fluids of opposite chirality(light blue and light grey regions) at filling factor $\nu=1$. A \textcolor{blue} {QD} is etched close to the inner edge of a fluid, enabling  electrons to tunnel between the edge and the dot, thus providing the resonant path in the Fano model.  \textcolor{blue}{QPC}1 splits the incoming electron beam from  $S_1$ or $S_2$ into two portions propagating along the upper and lower arms; \textcolor{blue} {QPC}2 recombines two portions back together, with resultant outgoing currents collected and measured at $D_1$ and $D_2$.  The two arms encircle a magnetic flux $\Phi$ and form a \textcolor{blue} {AB} ring.}  
\label{MZFI}
\end{figure}

We consider a four-terminal electronic Mach-Zehnder-Fano interferometer (\textcolor{Blue} {MZFI}), a device schematically sketched in Fig.~\ref{MZFI} that integrates a \textcolor{blue} {QD} into an electronic Mach-Zehnder interferometer (\textcolor{Blue} {MZI})\cite{Ji,Chung,Litvin,Neder1,Roulleau,Sukhorukov,Chalker,Levkivskyi,Bieri,Weisz,Neder2,Sueur,Tewari,Idrisov}.  An electronic \textcolor{blue} {MZFI} in a quantum Hall liquid at filling factor $\nu=1$ can be realized through the deposit of elaborately-tailored split metallic gates on the surface of the involved semiconductor heterostructure\cite{Ji,Neder1,Weisz}. Two inner edge states are employed as the chiral modes propagating along the two arms of the interferometer.  Ohmic contacts serve as carrier sources ($S_1,S_2$), each  injecting a carrier stream split at $x_1$ by the first quantum point contact (\textcolor{blue} {QPC}1)  to two partial beams. One moves along the upper arm with a nearby \textcolor{blue}{QD} at $x_d$, which provides the resonant path in the Fano model, while the other moves down the lower arm.  They recombine at  $x_2$ by the second quantum point contact (\textcolor{blue} {QPC}2)  and interfere, resulting in two complementary currents that are collected by another two ohmic contacts functioning as drains ($D_1,D_2$). 

The Hamiltonian describing such a device is $H_{D}=\sum_{\alpha=u,l}H_\alpha+H_d+H_{ul}+H_{ud}$. While $H_d$ and $H_{ud}$ explain the localized mode in the \textcolor{blue} {QD} and coupling between the \textcolor{blue}{QD} and the upper edge, $H_\alpha$ and $H_{ul}$ reflect the free motion along each edge and interedge electron tunneling. In terms of field operator annihilating an electron at position $x$ on the edge $\alpha$, $\hat{\psi}_\alpha(x)$, the Hamiltonian of free chiral modes is defined as follows: $H_\alpha=-i\hbar v_\text{F} \int dx \hat{\psi}^\dagger_\alpha (x) [\partial_x+ieA_\alpha (x)/\hbar c ] \hat{\psi}_\alpha(x)$, where $v_F$ is the Fermi velocity and $A_\alpha(x)$ is the magnetic vector potential experienced by chiral electrons on the edge $\alpha$ at position $x$. The interedge tunneling Hamiltonian is written as $H_{ul} = \sum_{j=1,2}[t_j \hat{\psi}^\dagger_u(x_j)\hat{\psi}_l(x_j)+H.c.]$, where the tunneling amplitude is $t_j=2\hbar v_\text{F} \gamma_j$ and $H.c.$ denotes Hermitian conjugate. This indicates that electrons can be reflected or transmitted at the positions $x_1$ and $x_2$ of the two \textcolor{blue} {QPC}s. The \textcolor{blue}{QD} consists of  non-interacting levels $\epsilon_i$ with localized modes $\xi_i(x)$. It is assumed that the level spacing of the \textcolor{blue}{QD} $\Delta\epsilon_i\propto 1/L_d$ is comparably large as long as the circumference $L_d$ of the dot is small enough. In this case, tunneling via only one discrete state is prominent if the energy $\epsilon_d$ of this state $\xi_d(x)$  is close to  the Fermi energy of the upper edge.  To describe tunneling between the \textcolor{blue}{QD} and the upper edge, we express the \textcolor{blue}{QD} field  in terms of  annihilation operators $\hat\psi_d(x)=\hat d  \xi_d(x)+\sum_{i} \hat{d}_i \xi_i(x)$. The effective tunneling Hamiltonian $H_{ud}$  is given by substituting the \textcolor{blue}{QD} field  in the  projected subspace $\xi_d(x)$ into the bare tunneling term $[ t_{du} \hat{\psi}^\dagger_u(x_d)\hat\psi_d(x_d)+ H.c.]$ : $H_{ud}=[t_d\hat{\psi}^\dagger_u(x_d)\hat d+ H.c.]$, where the tunneling amplitude $t_d=t_{ud}\xi(x_d)=\sqrt{2}\hbar v_\text{F}\gamma_d/\sqrt{a}$ with $a$ the characteristic length of the contact. In this projected subspace, the \textcolor{blue}{QD} may be modelled by the Hamiltonian $H_d=\epsilon_d \hat{d}^\dagger \hat{d}$.  Under the gauge transformation $\hat{\psi}_\alpha(x) \rightarrow e^{\frac{-ie}{\hbar c}\int^x_{x_0} dx A_\alpha(x)}\hat{\psi}_\alpha(x)$,  $A_\alpha(x)$-dependence is removed  from the free edge Hamiltonian, while  the tunneling amplitudes $t_j$ is replaced by $t_je^{\frac{ie}{\hbar c}[\int^{x_j}_{x_0} dx A_u(x)-\int^{x_j}_{x'_0} dx A_l(x)]}$.

%\begin{figure}[!t
%\centering
%\includegraphics[width=3in]{fig5}
%\subfloat[]{\includegraphics[width=0.53\columnwidth]{fig2a-diffcon.eps}}
%\subfloat[]{
% \includegraphics[width=0.53\columnwidth]{fig2b-curr.eps}}
%\caption{Differential conductance $dI/dV$ in units of $e^2/h$ (a, left panel) and current $I$ in units of $e\Gamma/h$ (b, right panel) at zero temperature are plotted as a function of the  dot level $\epsilon_d/\Gamma $  with varying magnetic flux $\Phi/\Phi_0$. The parameters are: $T_1=T_2=R_1=R_2=0.5,eV/\Gamma=5$.}
%\label{fig 2}
%\end{figure}

\begin{figure}
       \includegraphics[width=1\linewidth]{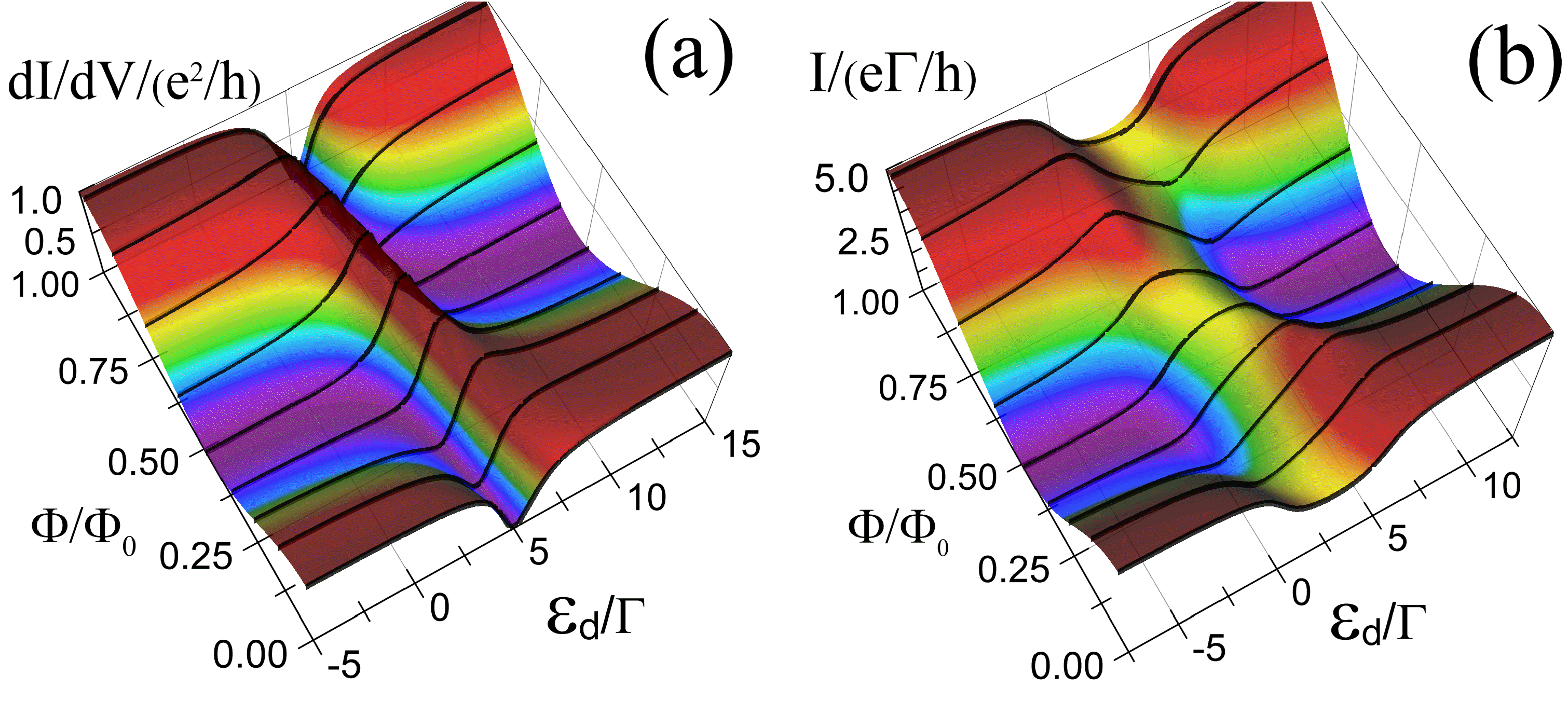}
    \caption{\label{fig:Fig2}Differential conductance $dI/dV$ in units of $e^2/h$ (a, left panel) and current $I$ in units of $e\Gamma/h$ (b, right panel) at zero temperature are plotted as functions of the  dot level $\epsilon_d/\Gamma $  with varying magnetic flux $\Phi/\Phi_0$. The parameters are: $T_1=T_2=0.5,eV=5\Gamma$.}
    \label{fig2}
\end{figure}

The \textcolor{blue}{MZFI} is assumed to be  coupled ballistically to four contacts at various electrochemical potentials $\mu_{i\alpha}$ via leads, where the upper or lower edge is denoted by $\alpha=u,l$ (or $1,2$) and the source or drain is labeled by $i=S,D$. 
Chiral electrons are assumed to propagate rightward from source to drain. The continuity equation $\partial_t \hat \rho^e_\alpha + \partial_x \hat I_\alpha =0(\alpha=u,l)$ may be used to read the current $I_\alpha(x)$ flowing along edge $\alpha$, where $\hat \rho^e_\alpha (x,t) =-e \hat{\psi}^\dagger_\alpha(x,t)\hat{\psi}_\alpha(x,t) $ is the chiral electron density operator on the edge $\alpha$.  The Heisenberg equation of motion for
$\hat \rho^e_\alpha (x,t)$ yields $\partial_t \hat \rho^e_\alpha (x,t)= [\hat \rho^e_\alpha (x,t), H_{D} \big ]/i\hbar  =-v_\text{F} \partial_x   \hat \rho^e_\alpha (x,t)$. Hence one finds 

\begin{equation}\label{current}
 \begin{split}
 I_\alpha(x) =& v_\text{F}\big <\rho^e_\alpha(x,t)\big >_{H_D}=ie\hbar v_\text{F} G^{<}_{\alpha \alpha}(x,t;x,t)\\
 &=\frac e h\int d\epsilon  i\hbar v_\text{F}  G^{<}_{\alpha \alpha}(x,x;\epsilon), 
 \end{split}
 \end{equation}
where  $G^{<}_{\alpha \alpha}(x,t;x',t')= i  \langle  \hat{\psi}^{\dagger} _{\alpha }(x',t')\hat{\psi}_\alpha(x,t)\rangle_{H_D}/\hbar$ is the lesser Green's function of the edge $\alpha$. $I_{\alpha}(x)$, when evaluated at a position $x$  in lead $D_i$,  represents the current  $I_{D_i}$ detected in drain $i$  following the notation $i=1$ if $\alpha=u$ and $i=2$ if $\alpha=l$, since the device is assumed to be connected ballistically to the leads.  
 
  We calculate the current $I_{D_1}$ transmitted into drain $1$, using the  equation-of-motion method in real space and the Keldysh technique (for details, See the Appendixes~\ref{App:A} and~\ref{App:B}), and find 
  \begin{equation}\label{uppercurrent}
   I_{D_1} = I_{D_1}^0+\frac e h \int^{+\infty}_{-\infty}d\epsilon [f_{S_2}(\epsilon)-f_{S_1}(\epsilon)]\mathcal{T}_{D_1S_2}(\epsilon), 
 \end{equation}
where 
\begin{equation}\label{directcurr}
I^0_{D_1}= \frac{e}{h}\int^{+\infty}_{-\infty}d\epsilon [f_{D_1}(\epsilon)-f_{S_1}(\epsilon)] 
\end{equation}
 is the direct current transmitted from $S_1$ into $D_1$ without interedge and edge-dot tunnelings, and 
 \begin{equation}\label{trand1s2}
  \mathcal{T}_{D_1S_2}(\epsilon)=|\sqrt{R_1T_2}+\sqrt{T_1R_2}\tau_d(\epsilon) e^{-i\phi}|^2 
  \end{equation}
    is the transmission probability from lead  $S_2$  to  lead  $D_1$.  Here $T_i=4\gamma^2_i/(1+\gamma^2_i)^2$ and $R_i=(1-\gamma^2_i)^2/(1+\gamma^2_i)^2$ are the transmission and reflection coefficients of the \textcolor{blue} {QPC}i, respectively; $\phi=\phi_{AB}+\phi_L$, $\phi_{AB}= 2\pi\frac{\Phi}{\Phi_0}$ is the \textcolor{blue} {AB} phase with $\Phi$ being the magnetic flux enclosed by two arms of the interferometer and $\Phi_0=hc/e$  the flux quantum, and $\phi_L=\epsilon \Delta L/\hbar v_F$ is the phase shift resulting from the length difference  $\Delta L=L_l-L_u$  of the two arms; $\tau_d(\epsilon)=\frac{\epsilon-\epsilon_d-i\Gamma}{\epsilon-\epsilon_d+i\Gamma}$ is the transmission amplitude through the position $x_d$ where the \textcolor{blue} {QD} is coupled, with $\Gamma=\hbar v_\text{F} \gamma^{2}_{d}/d $  the width of the resonant level $\epsilon_d$. It should be noted that, instead of $1$, the value of the heaviside function $\theta(x)$ at zero has been assumed to be $1/2$, i.e., $\theta(0)=1/2$ to obtain the transmission amplitude $\tau_d(\epsilon)$.   This assumption is supported within the scattering matrix approach in Appendix~\ref{App:C}. 

Eq.~\eqref{uppercurrent} shows that when both edges are coupled to the same source and drain only direct current persists, suggesting that quantization of the Hall conductance is independent of interedge tunneling and impurity scattering. 
For this reason, we will concentrate on the case in which $\mu_{S_1}=\mu_{D_1}=0, \mu_{S_2}=\mu_{D_2}=eV$  achieved by setting the electrochemical potentials of four contacts in this manner. The current incident from $S_2$, denoted by  $I=I_{D_1}=\frac e h \int d\epsilon [f(\epsilon-eV)-f(\epsilon)]\mathcal{T}_{D_1S_2}(\epsilon)$, is the only nonzero current.

\section{Analysis of Fano resonances}\label{Ana:Fano}

\subsection{Transmission}

  By defining a background transmission probability without edge-dot tunneling, $\mathcal{T}^0_{D_1S_2}(\phi)=|\sqrt{R_1T_2}+\sqrt{T_1R_2} e^{-i\phi}|^2$, and $\mathcal A=\sqrt{T_1R_1T_2R_2}$,  the transmission probability can be expressed in terms of a dimensionless detuning $\delta(\epsilon)=(\epsilon-\epsilon_d)/\Gamma$ as
\begin{align}\label{transsum}
\mathcal{T}_{D_1S_2}(\epsilon)=\mathcal{T}^0_{D_1S_2}(\phi)-4\mathcal A
  \frac{sin\phi\delta(\epsilon)+cos\phi}{\delta^2(\epsilon)+1}.
 \end{align}

Fano resonance is seen in the profile term of Eq.~\eqref{transsum}, where the antisymmetric part $\delta(\epsilon)/[\delta^2(\epsilon)+1]$ is represented by a peak at $\delta(\epsilon)=1$ and a dip at $\delta(\epsilon)=-1$, and the symmetric part $1/[\delta^2(\epsilon)+1]$ is characterized by a peak of width $2$ at $\delta(\epsilon)=0$. 
As $\phi=0$, this transmission probability exhibits a symmetric dip that is unique to the Fano resonance and fits neatly into the Fano profile.
\begin{equation}\label{transfano}
\mathcal{T}_{D_1S_2}(\epsilon)=\mathcal{T}^0_{D_1S_2}(\pi)+\frac{4\mathcal A}{1+q^2(\phi)}\frac{[\delta(\epsilon)+q(\phi)]^2}{\delta^2(\epsilon)+1}, 
\end{equation}
where the Fano asymmetry parameter is denoted by $q(\phi)=-tan(\phi/2)$. Eq.~\eqref{transfano} predicts one minimum at $\delta(\epsilon)=-q(\phi)$ and one maximum at $\delta(\epsilon)=1/q(\phi)$ in the transmission profile. If $q=0$, there appears a single symmetric peak; if $q\rightarrow \pm \infty$, there appears a single symmetric dip. In terms of the transmission phase $\phi_d(\epsilon)=arctan[2\delta(\epsilon)/(1-\delta^2(\epsilon))]-\pi$,  the transmission amplitude $\tau_d(\epsilon)$ admits an exponential expression $\tau_d(\epsilon)=e^{i\phi_d(\epsilon)}$. This leaves us with a well-known expression for the transmission probability $\mathcal T_{D_1S_2}=|\sqrt{R_1T_2}+\sqrt{T_1R_2}e^{i[\phi_d(\epsilon)-\phi]}|^2$. Constructive or destructive interference takes place when the phase difference $\phi_d(\epsilon)-\phi$ is $0$  or $\pi$.  The locations of the maxima and minima in the transmission spectrum can be found using the inverse trigonometric identities $2arctan(x)=arctan[2x/(1-x^2)],x^2<1$ and $arctan(x)+arccot(x)=\pi/2$. These relations are equivalent to the condition of constructive or destructive interference. If we additionally assume a symmetric interferometer, $\Delta L=0$, the Fano asymmetry parameter $q=-tan(\phi_{AB}/2)$ becomes merely a function of the \textcolor{blue}{AB} phase. Consequently, it is possible to fully tune Fano resonances in the transmission and linear conductance spectra of the \textcolor{blue} {MZFI}, simply by tuning  the asymmetry parameter with an external magnetic flux.

\subsection{Current and conductance} 

 At zero temperature,  the current is given by
\bea\label{currentexpression}
I&=&\frac{e\Gamma}{h}\Big[ \mathcal{T}^0_{D_1S_2}(\phi_{AB}) \frac{eV}{\Gamma}-4\mathcal A\big[\frac{sin\phi_{AB}}{2}ln(\delta^2(\epsilon)+1) \no \\
 & & + cos\phi_{AB}arctan\delta(\epsilon)\big]\Big|^{eV}_{0}\Big], 
 \eea
where $f(\epsilon)|^a_b=f(a)-f(b)$, and the differential conductance is calculated as 
\begin{equation}\label{diffcond}
\resizebox{0.9\hsize}{!}{$\begin{aligned}
\frac{dI}{dV}=\frac{e^2}{h}\Big[\mathcal{T}^0_{D_1S_2}(\phi_{AB})-4\mathcal A
  \frac{sin\phi_{AB}\delta(eV)+cos\phi_{AB}}{\delta^2(eV)+1}\Big] , 
\end{aligned}$}
\end{equation}
which can be also expressed as  $\frac{dI}{dV}=\frac{e^2}{h}\mathcal T_{D_1S_2}(eV)$.
 
Although their profiles appear to be quite different, the differential conductance and current are the same type of oscillatory function of the \textcolor{blue} {AB} phase with period $2\pi$ as the transmission probability, performing identical profile evolution as the \textcolor{blue} {AB} phase grows. However, since the current is the integration of the transmission probability over energy from $0$ to $eV$,  their profiles are closely related. and the current profile depends apparently on the applied voltage $V$. The profile function in Eq.~\eqref{currentexpression} possesses two unusual features: the arctangent term acts as the symmetric component, resulting in a peak at $\epsilon_d=eV/2$ that broadens with rising $eV$, and the logarithmic term behaves as the antisymmetric part, admitting a dip at $\epsilon_d=eV$ and a peak at $\epsilon_d=0$. Due to the cumulative effect of integration, the current profiles can be viewed as enlarged versions of the transmission ones.

 In actual experiments, it is more convenient to adjust the resonant level $\epsilon_d$  by varying the bias voltage to the  plunger gate of the dot, with the voltage $V$ being fixed.  
As functions of the dot level $\epsilon_d$ and the enclosed magnetic flux $\Phi/\Phi_0$, we display the differential conductance in Fig.~ \ref{fig2}\textcolor{blue}{(a)} and the current in Fig.~\ref{fig2}\textcolor{blue}{(b)} by setting $eV=5\Gamma$. These may be clearly seen by taking $T_1=T_2=0.5$. The differential conductance and the current, albeit having different profiles, exhibit fully tunable Fano resonances by adjusting the magnetic flux, as predicted. They also show identical profile progression over the increasing magnetic flux. In contrast to the sharp peak-dip structure in the differential conductance, the accumulation effect causes the peaks and dips in the current spectrum to broaden significantly and appear as arches and valleys, as can seen in Fig.~\ref{fig2}.  Observe that for any finite voltage $V$, the chiral transport is coherent, which is in line with the experimental observation\cite{Tewari} and the theoretical result\cite{Sukhorukov} in the study on docoherence in the electronic \textcolor{blue}{MZI}s. In our configuration, the \textcolor{blue}{QD} is  essentially an indispensable component of the coherent transport device as a whole. In contrast to the circumstances \cite{Weisz,Buks,Sprinzak} where the \textcolor{blue} {QD} functions as a detecting apparatus, it does not cause any dephasing.

\subsection{Shot noise}

Due to interference, Fano resonances arise in all the transmission probability through a \textcolor{blue} {MZFI}. Therefore one may expect Fano resonances in shot noise of  electronic \textcolor{blue} {MZFI}s as well.  At zero temperature, $\Delta I_{S1}=\Delta I_{S2}=0$, there is no noise in the incident stream according to the coherent scattering theory\cite{Buttiker1,Buttiker2,Blanter}. The currents flowing into the drain contacts, $I_{D1}$ and $I_{D2}$, are the only ones that fluctuate. Because of current conservation, $\Delta I_{D1}+\Delta I_{D2}=0$. The transmissions probabilities from lead $S_j$ to lead $D_i$ are associated with the scattering matrix elements of the system  by $\mathcal{T}_{D_iS_j}=|s_{D_iS_j}|^2$,  Flux conservation yields $\sum_{i=1,2}\mathcal{T}_{D_iS_j}=1$. As we show in Appendix~\ref{App:D}, nonzero components of the shot noise power tensor  are $S=S_{D_1D_1}=S_{D_2D_2}=-S_{D_1D_2}=-S_{D_2D_1}$. They are given by (for details, see
Appendix~\ref{App:D})  
\begin{equation}\label{shotnoise1}
S=\frac{e^2}{\pi \hbar}\int^{eV}_{0}d\epsilon \mathcal T_{D_1S_2}(\epsilon)\mathcal T_{D_2S_2}(\epsilon), 
\end{equation}
which gives differential shot noise 
\begin{widetext}
\begin{equation}\label{diffshotnoise}
\frac{dS}{dV}=\frac{2e^3}{h}\Big[\mathcal T^{0}_{D_1S_2}(\phi_{AB})\mathcal T^0_{D_2S_2}(\phi_{AB})+8\mathcal A^2\big[cos(2\phi_{AB}) \frac{2\delta^2(eV)}{[\delta^2(eV)+1]^2}+sin(2\phi_{AB})\frac{\delta(eV)[\delta^2(eV)-1]}{[\delta^2(eV)+1]^2}\big]\Big], 
\end{equation}
and shot noise 
\begin{equation}\label{shotnoise}
S=\frac{e^2\Gamma}{h}\Big[\mathcal T^{0}_{D_1S_2}(\phi_{AB})\mathcal T^0_{D_2S_2}(\phi_{AB})\frac{eV}{\Gamma}+8\mathcal A^2\big[cos(2\phi_{AB})P_S(\epsilon_d,eV)+
 sin(2\phi_{AB})P_A(\epsilon_d,eV)\big]\Big], 
\end{equation}
\end{widetext}
where $\mathcal T^0_{D_2S_2}$ is defined similar to $\mathcal T^{0}_{D_1S_2}$, $P_S(\epsilon_d,eV)=\big[arctan\delta(\epsilon)
-\delta(\epsilon)/[\delta^2(\epsilon)+1]\big]\big|^{eV}_0$ and $P_A(\epsilon_d,eV)=\big[1/[\delta^2(\epsilon)+1]+ ln[\delta^2(\epsilon)+1]/2 \big]\big|^{eV}_{0}$
are the symmetric and antisymmetric parts of the profile function of the shot noise.  As functions of \textcolor{blue} {AB} phase, the differential shot noise and shot noise execute the identical oscillation with halved period $\pi$. The ratio of the  shot noise to the Poisson noise, $F=S/S_P=S/2eI$, is known as the Fano factor\cite{Blanter}. An equivalent definition of the differential Fano factor may be $F_d=(dS/dV)/2e(dI/dV)$. If the \textcolor{blue} {QD} is removed from the device, these two factors coincide and reduce to $F^0=F^0_d=\mathcal T^0_{D_2S_2}=|\sqrt{R_1R_2}-\sqrt{T_1T_2}e^{-i\phi_{AB}} |^2$, 
which is always no greater than $1$ and exhibit \textcolor{blue} {AB} oscillations. Once the \textcolor{blue} {QD} is involved, the Fano factor $F$ evolves in a complicated way, while the differential Fano factor $F_d$  remains rather simple and acquires the form of the transmission  $\mathcal T_{D_2S_2}(eV)$, thus showing perfect Fano resonances.

\begin{figure}
      \includegraphics[width=1\linewidth]{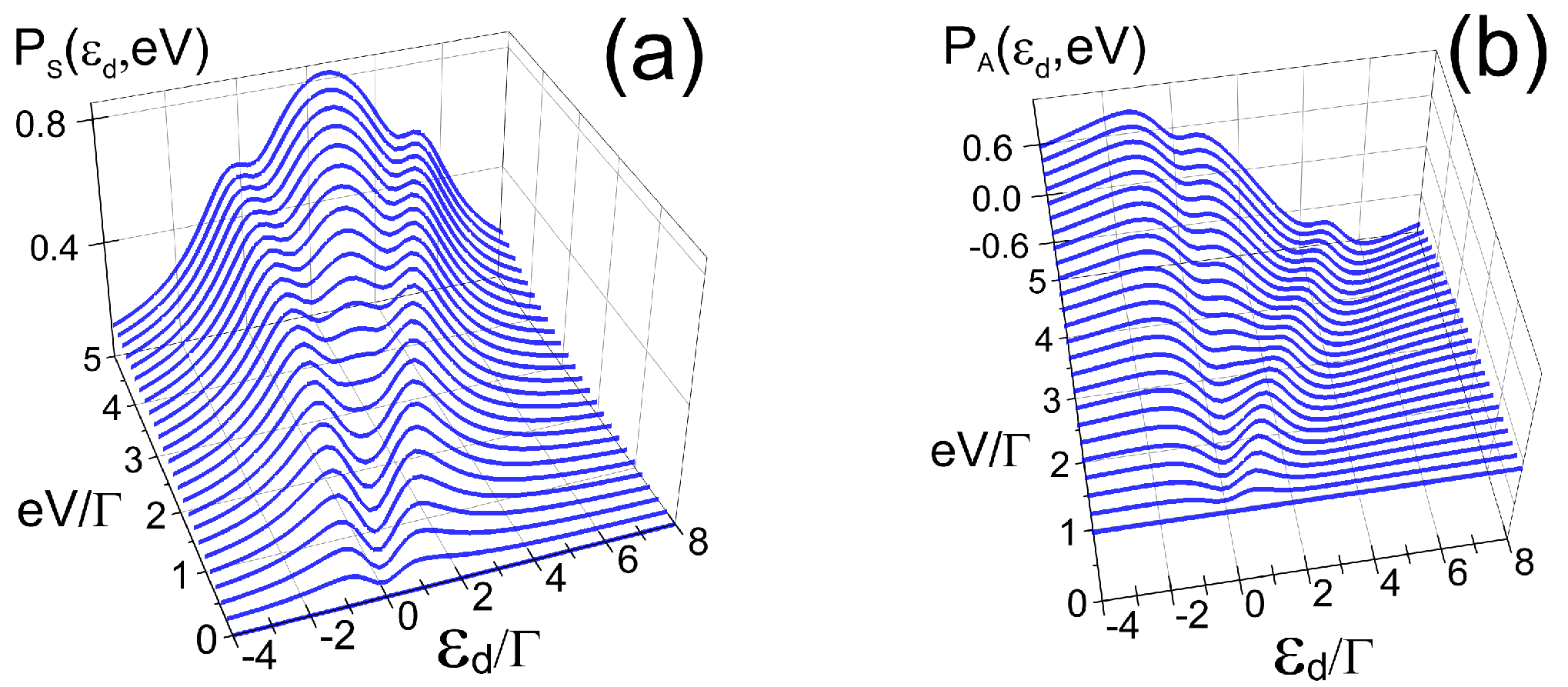}
    \caption{Evolution of symmetric $P_S(\epsilon,eV)$ (a, left panel) and antisymmetric $P_A(\epsilon,eV)$ (b, right panel) parts of the profile function in shot noise as the voltage $eV$ increases from $0$ to $5\Gamma$.}
    \label{fig3}
 \end{figure}

\begin{figure}
     \includegraphics[width=1\linewidth]{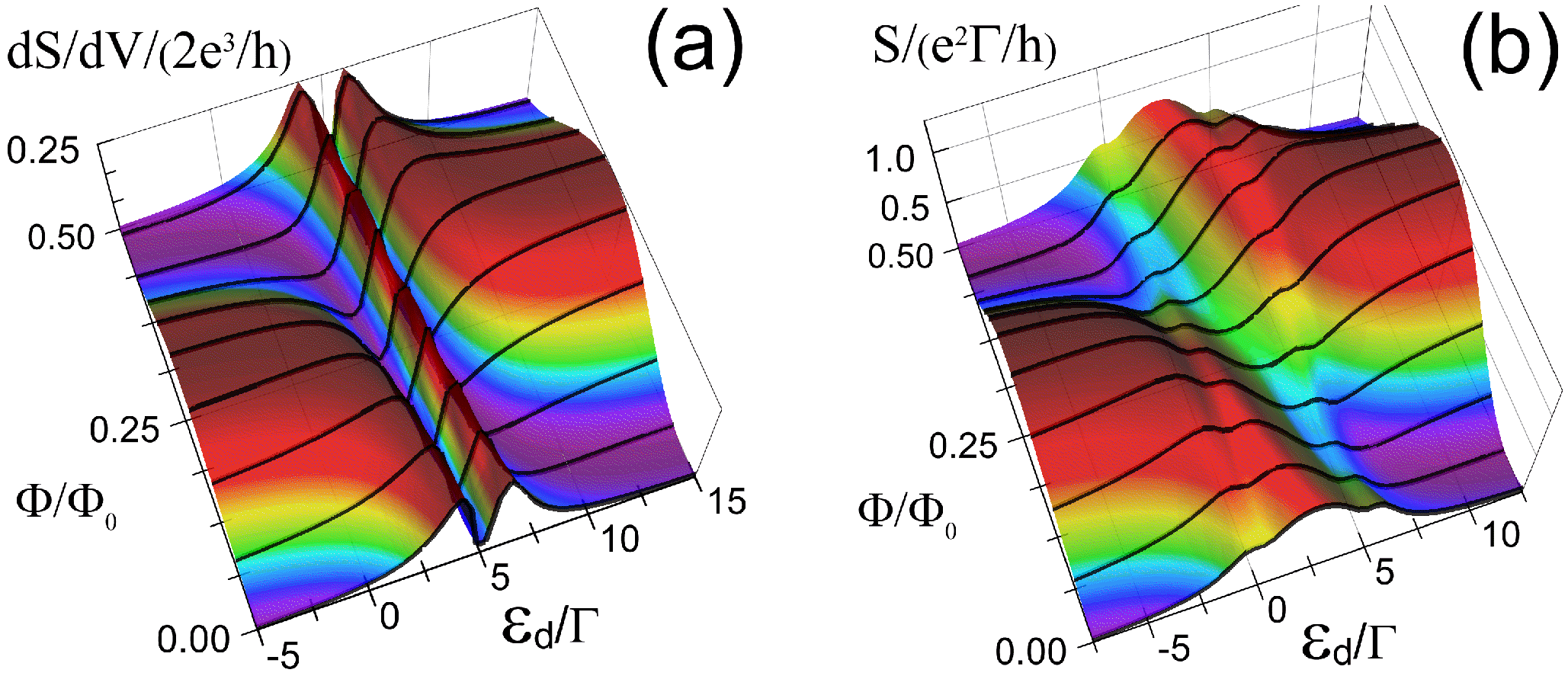}
    \caption{Differential shot noise $dS/dV$ in units of $2e^3/h$ (a, left panel) and shot noise $S$ in units of $e^2\Gamma/h$ (b, right panel)  are plotted as functions of the dot level $\epsilon_d/\Gamma $  with varying magnetic flux $\Phi/\Phi_0$. The parameters are the same as in Fig. \ref{fig2}}
    \label{fig4}
 \end{figure}

The product of  two transmission probabilities, $\mathcal T_{D_1S_2}(eV)$ and $\mathcal T_{D_2S_2}(eV)$, determines the differential shot noise. It takes maximum values when one of them is halfway between its maximum and minimum, and minimum values when either of the two coefficients is maximal or minimal. As a result, the differential shot noise spectrum  shows dips where the differential conductance appears as a peak or a dip, and peaks halfway between the ridge and the valley in the differential conductance spectrum. Such an intuitive picture can be verified by examinating the profile function of differential shot noise in Eq.~\eqref{diffshotnoise}, which is reminiscent of the line shape of complex susceptibility of three-level atomic systems in electromagnetically induced transparency(\textcolor{blue}{EIT}) phenomenon\cite{Scully}. \textcolor{blue}{EIT}, or absorption cancellation, is caused by destructive interference between the two coherent routes for absorption and is comparable to Fano antiresonance. The antisymmetric part of the profile in the sine term admits, on the positive side of the $\delta(eV)$-axis, a sharp dip of depth $0.25$ at $\delta(eV)=\sqrt{2}-1$ and a broad peak of height $0.25$ at $\delta(eV)=1+\sqrt{2}$. The symmetric part of the profile in the cosine term is peaked symmetrically at $\delta(eV)=\pm 1$ and drops to zero at the midpoint $\delta(eV)=0$. Depending on $\phi_{AB}$, either of these two terms dominates, or both cooperate, resulting in symmetric double-peak or double-dip profiles, or asymmetric profiles with a deformed double-peak or double-dip structure in the differential shot noise spectrum.
In the case of shot noise, the symmetric component $P_S(\epsilon,eV)$ has a profile of two or three peaks symmetric about the $\epsilon_d=eV/2$, depending on the ratio of $eV/\Gamma$. On the other hand, the antisymmetric component $P_A(\epsilon,eV)$ gives us a profile of a sharp peak with a broad dip on the right and a broad peak on the left as long as $eV/<2\Gamma$. The sharp peak splits and gradually develops into two side shoulders that are located at $\epsilon_d=0$ and $\epsilon_d=eV$ as $eV$ increaces. Figures \ref{fig3}\textcolor{blue}{(a)} and \ref{fig3}\textcolor{blue}{(b)} display clearly the evolutions of the symmetric and antisymmetric components of the shot noise profile with increasing voltage $eV$. We stress that, unique to Fano resonances, a profile of symmetric double-peak with a dip at the center $\epsilon_d=eV/2$ and a profile of symmetric double-peak or triple-peak should be detected in differential shot noise and shot noise spectra of a \textcolor{blue}{MZFI}.  We provide profile evolutions of the differential shot noise from Eq.~\eqref{diffshotnoise} in Fig.~\ref{fig4}\textcolor{blue}{(a)} and  shot noise from Eq.~\eqref{shotnoise} in Fig.~\ref{fig4}\textcolor{blue}{(b)}, where the magnetic flux $\Phi/\Phi_0$ varies from $0$ to $0.5$. Figures~\ref{fig4}\textcolor{blue}{(a)} and~\ref{fig4}\textcolor{blue}{(b)} confirm  our analysis.  Furthermore, the evolution pattern in both the differential shot noise and shot noise is identical to that in the differential conductance and current throughout one evolution cycle. Profile evolutions of all the transport spectra of a \textcolor{blue}{MZFI} are determined by sine and cosine functions, therefore, this result comes as no surprise.

\section{Discussion}\label{Dis}

\subsection{Asymmetric \textcolor{blue}{MZFI}}

We have investigated Fano resonances in transport spectra for a symmetric \textcolor{blue}{MZFI} with $\Delta L=0$.  The presence of interferometer arm asymmetry, for example, $\Delta L=L_l-L_u> 0$, will introduce another energy scale $E_A=\hbar v_F/\Delta L$ in the problem. The replacement $\phi_{AB}\rightarrow \phi_{AB}+eV/E_A$ can be used to explain how the asymmetry in interferometer arm lengths affects differential transport spectra including differential conductance, differential shot noise, and differential Fano factor. 
As for the asymmetry effect on the integrated spectra such as the current and shot noise, we are interested in the case where there is a small mismatch between the arm lengths such that $E_A \gg eV, \epsilon_d,\Gamma$. The presence of the asymmetry effect is indicated by a slight modification of the symmetric ($P_S$) and antisymmetric ($P_A$) parts of the profile functions, so that, up to first order in $1/E_A$, $P'_S\approx (1-\frac{\Gamma}{E_A}) P_S+\frac{\epsilon_d}{E_A}P_A, P'_A\approx (1-\frac{\Gamma}{E_A})P_A-\frac{\epsilon_d}{E_A}P_S$ for current, and $P'_S\approx (1-\frac{2\Gamma}{E_A}) P_S+\frac{2\epsilon_d}{E_A}P_A-\frac{2\Gamma}{E_A}arctan\delta(\epsilon)\big|^{eV}_0, P'_A\approx (1-\frac{2\Gamma}{E_A})P_A-\frac{2\epsilon_d}{E_A}P_S-\frac{\Gamma}{E_A}ln[\delta^2(\epsilon)+1]\big|^{eV}_{0}$ for  shot noise. The integrations in the current and shot noise equations cannot be further simplified but can be directly computed numerically if the mismatch $\Delta L$ is comparable with $eV$. In this instance, the shot noise and current profiles will be distorted, but the profile evolution will remain intact.

\subsection{\textcolor{blue}{QD} in the Coulomb blockade or Kondo regime}

When taking into account the Coulomb interaction $U$ between electrons in the \textcolor{blue}{QD}, one may question if Fano resonances can be observed in transport spectra of a \textcolor{blue}{MZFI}. We believe that the formalism is applicable to an interacting resonant level that can be described by a single level Anderson model and thus Fano resonances persist even if an interacting \textcolor{blue}{QD} is involved.  If not considering possible quantum phase transitions, chiral electrons always go through the point where the \textcolor{blue} {QD} is side-coupled. This imposes a strict constraint on the formalism. In the case of a noninteracting \textcolor{blue}{QD}, the bare 
propagator along the upper edge is $g^r_{uu}(x,x')=g^{0r}_{uu}(x,x')\tau_d(\epsilon)$, $\tau_d(\epsilon)=1-2i\Gamma g^r_{dd}=\frac{\epsilon-\epsilon_d-i\Gamma}{\epsilon-\epsilon_d+i\Gamma}$  provided $g^r_{dd}(\epsilon)=[(g^{0r}_{dd}(\epsilon))^{-1}-\Sigma_d(\epsilon)]^{-1}=(\epsilon-\epsilon_d+i\Gamma)^{-1}$. The details are given in
Appendix~\ref{App:A(4)}. If the Coulomb interaction is present, one must resort to approximate schemes to tackle this problem.  Modifying the Green's function of the isolated  \textcolor{blue}{QD} is a clumsy but sensible solution under the constraint $Im \Sigma_d(\epsilon)=-i\Gamma$.
One can choose $Re\Sigma_d(\epsilon)=0 $ and $g^{0r}_{dd}(\epsilon)= (1-n_d)(\epsilon-\epsilon_d+i0^+)^{-1}+n_d(\epsilon-\epsilon_d-U+i0^+)^{-1}$ in the Coulomb blockade regime\cite{Meir,Haug}, $g^{0r}_{dd}(\epsilon)=(\epsilon-\epsilon_d+i0^+)^{-1}$ and $\epsilon-\epsilon_d-Re\Sigma_d(\epsilon)=(\epsilon-\alpha)\Gamma/(2k_BT_K)$) in the Kondo regime\cite{Houghton,Madhavan}. Here, $T_K$ is the Kondo temperature, $\alpha$ is a constant, and $n_d$ is the occupation number of the resonant level $\epsilon_d$. 
Thus, one may convince himself that even in the case when the \textcolor{blue} {QD} is in the equilibrium  the Coulomb blockade regime or Kondo regime ,fully tunable Fano resonances could be expected.

  \section{Concluding remarks}\label{Con}
  \label{Conrem}
  
  To summarize, our results indicate that Fano resonances in chiral electron transport require an extra direct path to emerge. Motivated by this finding, we suggest an electronic \textcolor{blue} {MZFI}, which is possibly the simplest and cleanest setting to explore the Fano effect in chiral electronic transport. Despite their simplicity, fully tunable Fano resonances can be easily achieved by altering an external magnetic field in almost all transport spectra of this simple arrangement. They are robust with respect to device parameters such as interedge couplings, edge-\textcolor{blue}{QD} coupling, and arm lengths of the interferometer. Therefore, parameters to fit Fano profiles of transport spectra are not needed in chiral electronic transport. Future directions would be of great interest in adapt it to a more intricate edge structure such as the edge modes in the fractional quantum hall systems and studying the Fano effect in transport through an asymmetric \textcolor{blue}{MZFI} under decoherence. Another possibility could be developing more accurate analytical and numerical approaches to examine the overlapped Fano resonances with a multi-level \textcolor{blue} {QD} and the Fano effect with a strongly correlated \textcolor{blue} {QD}.

  %%%%%%%%%%%%%%%%%%%%%%%%%%%%%%%%%%%%%%%%%%%%%%%%%%
\section*{Acknowledgments}
This work is supported by National Natural Science Foundation of China (Grant No. 12464019) and Science and Techology Research Project of Jangxi Provincial Department of Education (Grant No. GJJ2200324). 

%%%%%%%%%%%%%%%%%%%%%%%%%%%%%%%%%%%%%%%%%

%%%%%%%%%%%%%%%%%%%%%%%%%%%%%%%%%%%%%%%%%%%%%%%%%%

\appendix

\section{Derivation of the current Eq.~\eqref{uppercurrent} in the absence of contact Hamiltonian} \label{App:A} 

In this Appendix, we present a detailed derivation of the current~\eqref{uppercurrent} without considering the contact Hamiltonian, based on the equation-of-motion method and the Keldysh 
technique. Sec.~\ref{App:A(1)} shows the details of how to obtain Dyson's equations for the various full retarded Green's functions from their equations of motion. The Keldysh equation for the lesser Green's function of the upper edge is given in Sec.~\ref{App:A(2)}. Sec.~\ref{App:A(3)} gives the expressions for the various bare Green's functions of the isolated edges in the absence of a \textcolor{blue}{QD}. Modified retarded Green's functions of the isolated upper edge with a side-attached \textcolor{blue}{QD} are presented in Sec.~\ref{App:A(4)}.  In Sec.~\ref{App:A(5)} we calculate the various full retarded Green's functions and present their explicit expressions.  Finally we derive Eq.~\eqref{uppercurrent} in Sec.~\ref{App:A(6)}. 

\subsection{Dyson's equations for the full retarded Green's functions of the edges}\label{App:A(1)}

In this section we give a detailed derivation of Dyson's equations for the full propagators (retarded Green's functions) $G^{r}_{\alpha\beta}(x,t;x',t')$ ($ \alpha,\beta=u,l$) with the equation-of-motion method.  The full propagators are defined as
\begin{align}
 G^{r}_{\alpha\beta}(x,t;x',t')=-\frac i\hbar \theta(t-t')\big \langle \big \{\hat{\psi}_\alpha(x,t),\hat{\psi}^\dagger_\beta(x',t') \big\}\big\rangle_{H_{D}}. \no
 \end{align}
 
As an example, we first consider the equation of motion for the  propagator $G^{r}_{uu}(x,t;x',t')$ of the upper edge
\begin{widetext}
 \begin{equation}\label{eqsA1}
 i\hbar \partial_t  G^{r}_{uu}(x,t;x',t')=-i\hbar v_\text{F}\partial_xG^{r}_{uu}(x,t;x',t')+ 
  \sum_{i=1,2}t_iG^r_{lu}(x_i,t;x',t')\delta(x-x_i)+\delta(t-t')\delta(x-x'),
 \end{equation}
 \end{widetext}
  where 
 \begin{align}
  G^r_{lu}(x_i,t;x',t')=-\frac i\hbar \theta(t-t') \big <\big \{\hat{\psi}_l(x_i,t),\hat{\psi}^\dagger_u(x',t') \big\}\big>_{H_{D}},\no
  \end{align}
 are the mixed propagators associated with both the upper and lower edges.
 
 The  retarded Green's functions of the isolated upper or lower edge are denoted by  
 \bea
 g^{0r}_{\alpha\alpha}(x,x';t)=-\frac i\hbar\theta(t-t') \big <\big \{\hat{\psi}_\alpha(x,t),\hat{\psi}^\dagger_\alpha(x',t') \big\}\big>_{H_\alpha}, \no
 \eea
  To keep things simple, we take a decoupling scheme. First we consider the equation of motion for the propagators of the upper edge which is decoupled from the \textcolor{blue}{QD}. The bare Green's functions of the isolated upper edge is denoted by $g^0_{uu}$, and the Green's functions of the upper edge coupled to the \textcolor{blue}{QD} but decoupled from the lower edge is denoted by $g_{uu}$,  which can be easily obtained from the equation of motion without considering interedge tunneling. We leave this task in Sec.~\ref{App:A(4)}.
   
   The equation of motion for $g^{0r}_{uu}(x,x';t)$ reads  
 \bea 
 \begin{split}
i\hbar \partial_t g^{0r}_{uu}(x,t;x',t')&=\delta(t-t')\delta(x-x')+ \theta(t) \\
&\big <\big \{\big[\hat{\psi}_u(x,t),  H_\alpha(x,t)\big],\hat{\psi}^\dagger_u(x',t')\big \}\big>_{H_\alpha},\no
\end{split}
 \eea
 which can be simplified as
 \begin{equation}\label{eqsA2}
 i\hbar (\partial_t+v_\text{F}\partial_x) g^{0r}_{uu}(x,t;x',t')=\delta(t-t')\delta(x-x').
 \end{equation} 
 With this equation,  Eq.~\eqref{eqsA1} can be cast in the following form
\bea
G^{r}_{uu}(x,t;x',t')&=&g^{0r}_{uu}(x,t;x',t')+\!\int\! d\tau \!\sum_{i=1,2} g^{0r}_{uu}(x,t; x_i,\tau)\no\\ 
  & & t_i G^r_{lu}(x_i,\tau;x',t'),
  \eea
which is Fourier transformed to 
\bea \label{eqsA4}
 G^{r}_{uu}(x,x';\epsilon)&=&g^{0r}_{uu}(x,x';\epsilon) +\sum_{i=1,2} g^{0r}_{uu}(x, x_i,\epsilon) \no  \\ && t_iG^r_{lu}(x_i,x';\epsilon).
 \eea

Following the same procedure, one gets the  equation for the mixed Green's functions $G^r_{lu}(x,x';\epsilon)$ 
\bea\label{eqsA5}
\hspace{-0.2cm}
 G^{r}_{lu}(x,x';\epsilon)=\sum_{j=1,2} g^{0r}_{ll}(x,x_j,\epsilon)
 t_j^* G^r_{uu}(x_j,x';\epsilon). 
\eea

Substitution of Eq.~\eqref{eqsA5} into Eq.~\eqref{eqsA4} yields a standard Dyson's equation
\bea\label{eqsA6}
 G^{r}_{uu}(x,x';\epsilon)&=&g^{0r}_{uu}(x,x';\epsilon)+\sum_{i,j=1,2} g^{0r}_{uu}(x, x_i,\epsilon) \no\\&&
 \Sigma^{r}_{u}(x_i,x_j;\epsilon)  G^r_{uu}(x_j,x';\epsilon). 
 \eea
Here we have defined the retarded self-energies  $\Sigma^{r}_{u}(x_i,x_j;\epsilon)=t_it_j^*g^{0r}_{ll}(x_i, x_j,\epsilon)$ due to interedge tunneling.

If we set $x=x_i,x'=x_j$, Eq.~\eqref{eqsA6} is specified as
\bea\label{eqsA7}
 G^{r}_{uu}(x_i,x_j;\epsilon)&=&g^{0r}_{uu}(x_i,x_j;\epsilon)+\sum_{m,n=1,2} g^{0r}_{uu}(x_i, x_m,\epsilon)\no \\
 && \Sigma^{r}_{u}(x_m,x_n;\epsilon)    G^r_{uu}(x_n,x_j;\epsilon). 
 \eea
Referring the terms associated with arguments $x_i$ and $x_j$, $i,j=1,2$,  to the elements of $2\times 2$ matrices, we can rewrite the above equation  in matrix form  
\begin{align}
\hat{G}^{r}_{uu}(\epsilon)=\hat{g}^{0r}_{uu}(\epsilon)+\hat{g}^{0r}_{uu}(\epsilon) \hat{\Sigma}^{r}_{u}(\epsilon)\hat{G}^{r}_{uu}(\epsilon).\no
\end{align}
From now on we suppress the argument $\epsilon$ for easy representation and recover it when necessary. Solving for $G^r_{uu}(x_i,x_j)$ in terms of $\hat g^{0r}_{uu}$ gives
\begin{align}\label{eqsA8}
G^{r}_{uu}(x_i,x_j)&=&\sum_k [(\hat{1}-\hat{g}^0_{uu} \hat{\Sigma} ^{r}_{u})^{-1}]_{ik} g^{0r}_{uu}(x_k,x_j)\no \\
&=&\sum_k g^{0r}_{uu}(x_i,x_k)[(\hat{1}-\hat{\Sigma} ^{r}_{u}
\hat{g}^0_{uu} )^{-1}]_{kj}. 
\end{align}

Setting $x=x_i$, we proceed by iteration 
\bea \label {eqsA9}
G^{r}_{uu}(x_i,x')&=&g^{0r}_{uu}(x_i,x')+\sum_{jk}  g^{0r}_{uu}(x_i,x_j) \Sigma^{r}_{u}(x_j,x_k) \no\\ 
&&g^{0r}_{uu}(x_k,x') 
  +\sum_{jkmn}  g^{0r}_{uu}(x_i,x_j)\Sigma^{r}_{u}(x_j,x_k)  \no   \\
  && g^{0r}_{uu}(x_k,x_m)\Sigma^{r}_{u}(x_m,x_n)g^{0r}_{uu}(x_n,x')\!\cdots. 
\eea

It is easy to convince oneself that the infinite order iteration results in
\bea\label{eqsA10}
\hspace{-0.3cm} 
G^{r}_{uu}(x_i,x')=\sum_{j} [(\hat{1}-\hat{g}^{0r}_{uu}\hat{\Sigma}^{r}_{u})^{-1}]_{ij} g^{0r}_{uu}(x_j,x').
\eea
Analogously one has  
\bea\label{eqsA11}  
\hspace{-0.2cm}
G^{r}_{uu}(x,x_j)=\sum_{i} g^{0r}_{uu}(x,x_i) [(\hat{1}-\hat{\Sigma}^{r}_{u} \hat{g}^{0r}_{uu})^{-1}]_{ij}.  
\eea

Substitution of  Eq.~\eqref{eqsA10} into Eq.~\eqref{eqsA6} allows us to express the  propagators of the upper edge  in terms of the bare ones
\bea \label{eqsA12}
 G^{r}_{uu}(x,x';\epsilon)&=&g^{0r}_{uu}(x,x';\epsilon)+\sum_{i,j} g^{0r}_{uu}(x, x_i,\epsilon)\slashed{\Sigma}^{r}_{u}(x_i,x_j,\epsilon)\no \\
   &&
              g^{0r}_{uu}(x_j,x';\epsilon),              
 \eea
where we have introduced an improper retarded self-energy matrix $ \hat{\slashed \Sigma}^{r}_{u}=\hat \Sigma^{r}_{u}  (\hat{1}-\hat{g}^{0r}_{uu}\hat{\Sigma}^{r}_{u})^{-1}$.

The propagators of the lower edge can be obtained in a similar manner 
 
 \bea \label{eqsA13}
 G^{r}_{ll}(x,x';\epsilon)&=&g^{0r}_{ll}(x,x';\epsilon)+\sum_{i,j} g^{0r}_{ll}(x, x_i,\epsilon)\slashed\Sigma^{r}_{l}(x_i,x_j;\epsilon)
             \no\\&& g^{r}_{ll}(x_j,x';\epsilon),              
 \eea
where $\slashed \Sigma^{r}_{l}=\Sigma^{r}_{l} (\hat{1}-\hat{g}^{0r}_{ll}\hat{\Sigma}^{r}_{l})^{-1}$ is the improper retarded self-energy associated with the retarded Green's functions of the lower edge, and $\Sigma^{r}_{l}(x_i,x_j;\epsilon)=t_i^*t_jg^{0r}_{uu}(x_i, x_j,\epsilon)$ are the corresponding retarded self-energies.

Inserting Eq.~\eqref{eqsA10} into Eq.~\eqref{eqsA5}, we get the expression for the mixed  propagators  $G^{r}_{lu}(x,x';\epsilon)$ from the upper edge to the lower edge
\bea\label{eqsA14}
G^{r}_{lu}(x,x';\epsilon)&=&\sum_{i,j=1,2} g^{0r}_{ll}(x,x_i,\epsilon)t_i^* [(\hat{1}-\hat{g}^{0r}_{uu}\hat{\Sigma}^{r}_{u})^{-1}]_{ij} \no \\ && g^{0r}_{uu}(x_j,x',\epsilon). 
\eea
The mixed  propagators  $G^{r}_{ul}(x,x';\epsilon)$ from the lower edge to the upper edge are derived as
\bea\label{eqsA15}
G^{r}_{ul}(x,x';\epsilon)&=&\sum_{i,j=1,2} g^{0r}_{uu}(x,x_i,\epsilon)t_i [(\hat{1}-\hat{g}^{0r}_{ll}\hat{\Sigma}^{r}_{l})^{-1}]_{ij} \no\\&& g^{0r}_{ll}(x_j,x',\epsilon). 
\eea

\subsection{The lesser Green's functions of the upper edge}\label{App:A(2)}

 The strategy to find a lesser Green's function $G^<$ is to apply the Langreth rules\cite{Haug} to the integral equation for the retarded one. Infinite order iteration of the resultant integral equation will lead to the keldysh equation for the lesser Green's function.  Starting with  Eq.~\eqref{eqsA6} for the retarded Green's function $G^r_{uu}(x,x';\epsilon)$,  We formulate it  in algebraic form
\bea \label{eqsA16}
G^r=g^r+g^r \Sigma^rG^r, 
\eea
where we suppressed the subscripts $uu$ for simplicity and recover them when necessary.  Applying the Langreth rule $(AB)^<=A^rB^<+A^<B^a$ to Eq.~\eqref{eqsA16}, we obtain the  equation for the lesser Green's function $G^<$
\bea\label {eqsA17}
G^<=g^<+g^<\Sigma^a G^a+g^r \Sigma^<G^a+g^r \Sigma ^r G^<. 
\eea
This equation can be iterated, and the infinite order iteration results in the well-known Keldysh equation
\bea\label{eqsA18}
G^<=(1+G^r \Sigma^r)g^<(1+ \Sigma^aG^a)+G^r \Sigma^< G^a. 
\eea 
Using the relations $g^<=f(\epsilon)(g^a-g^r)$, $1+G^r \Sigma^r=G^r (g^r)^{-1}$ and  $1+\Sigma^a G^a=(g^a)^{-1}G^a$, the keldysh equation ~\eqref{eqsA18} is transformed into
\bea \label{eqsA19}
G^<=f(\epsilon)(G^a-G^r)+\hat{G}^rf(\epsilon)(\Sigma^r-\Sigma^a)G^a+G^r \Sigma^<G^a. \no\\ 
\eea

For chiral electrons, the retarded (or advanced) Green's function propagating from $x$ back to $x$ is the same as its isolated counterpart, i.e.,  $G^{r/a}_{uu}(x,x;\epsilon)=g^{0r/a}_{uu}(x,x;\epsilon)$. As a consequence, the above Keldysh equation can be further simplified into 
\bea\label{eqsA20}
G^{<}_{uu}(x,x)&=&[f_{S_1}(\epsilon)-f_{eq}(\epsilon)][g^{a}_{uu}(x,x)-g^{r}_{uu}(x,x)]+\no\\
&&\sum_{i,j}G^{r}_{uu}(x,x_i)\Sigma^{<}_{u,ij}G^{a}_{uu}(x_j,x).
\eea
Here we have defined a lesser self-energy matrix $\hat \Sigma^<_{u}$  
\begin{equation}\label{eqsA21}
\Sigma^<_{u,ij}=\Sigma^<_u(x_i,x_j)=i[f_{S_2}(\epsilon)-f_{S_1}(\epsilon)]\Gamma^u(x_i,x_j), 
\end{equation}
with coupling matrix elements given by
\begin{equation}\label{eqsA22}
\Gamma^u(x_i,x_j)=i[\Sigma^{r}_{u}(x_i,x_j)-\Sigma^{a}_{u}(x_i,x_j)]. 
\end{equation}

Notice that  an equilibrium term $f_{eq}(g^a-g^r)$ in Eq.~\eqref{eqsA20} has been discarded, since it does not contribute to current.

\subsection{ Various bare Green's functions of the isolated edges}\label{App:A(3)}

In order to find the various full Green's functions, one needs to know the bare ones of the isolated edges. We first derive these functions in the absence of a \textcolor{blue}{QD}. The mode expansion of chiral field operators in the thermodynamical limit may be written as  
\begin{align}\label{eqsA23}
\hat{\psi}_\alpha(x,t)=\int^{+\infty}_{-\infty} \frac{dk}{2\pi} \hat c_\alpha(k) e^{ik(x- v_\text{F}t)},\alpha=u,l 
\end{align}
where $\hat c_\alpha(k)$ is  the annihilation operator for a state with wave number $k$ on the edge $\alpha$. Then one arrives at
\bea
g^{0r}_{\alpha\alpha}(x,t;x',t')&=&\!-\frac i\hbar \theta(t-t') \big <\big \{\hat{\psi}_\alpha(x,t),\hat{\psi}^\dagger_\alpha(x',t') \big\}\big>_0\no \\&=&\!-\frac i\hbar \theta(t-t')\! \int^{+\infty}_{-\infty}\! \frac{dk}{2\pi} e^{-iv_\text{F}k(t-t')} e^{ik(x-x')}, \no
\eea
Fourier transform of which gives
\bea\label{eqsA24}
g^{0r}_{\alpha\alpha}(x,x';\epsilon)=-\frac{i}{\hbar v_\text{F}}\theta(x-x') e^{i\epsilon(x-x')/\hbar v_\text{F}}. 
\eea

The corresponding advanced and lesser Green's functions can be calculated by direct substitution
\bea
g^{0a}_{\alpha\alpha}(x,x';\epsilon)&=&\frac{i}{\hbar v_\text{F}}\theta(x'-x) e^{i\epsilon(x-x')/\hbar v_\text{F}}, \\ \label{eqsA25}
g^{0<}_{\alpha\alpha}(x,x';\epsilon)&=&\frac{i}{\hbar v_\text{F}}f_{S_\alpha}(\epsilon) e^{i\epsilon(x-x')/\hbar v_\text{F}},  \label{eqsA26}
\eea
where $f_{S_\alpha}(\epsilon)=f(\epsilon-\mu_{S_\alpha})$ is the Fermi-Dirac distribution function of the source $S_\alpha$, which injects electrons in the edge, $\alpha=u,l$. In our setup, $S_u=S_1, S_l=S_2$ and $D_u=D_1,D_l=D_2$. For this reason, $\alpha=u,l$ and $i=1,2$ are adopted interchangeably  to label the contacts. 

Since
\bea\label{eqsA27}
\hspace{-0.3cm}
g^{0a}_{\alpha\alpha}(x,x';\epsilon)-g^{0r}_{\alpha\alpha}(x,x';\epsilon)=\frac{i}{\hbar v_\text{F}} e^{i\epsilon(x-x')/\hbar v_\text{F}},
\eea
the lesser Green's functions can be rewritten in a desired form
\bea \label{eqsA28}
\hspace{-0.3cm}
g^{0<}_{\alpha\alpha}(x,x';\epsilon)=f_{S_\alpha}[g^{0a}_{\alpha\alpha}(x,x';\epsilon)-g^{0r}_{\alpha\alpha}(x,x';\epsilon)].  
\eea

To apply Eq.~\eqref{eqsA28} to non-equilibrium transport problems, it will be proved to be useful to make the following decomposition
since the current depends only on the difference of distribution functions
\bea \label{eqsA29}
g^{0<}_{\alpha\alpha}(x,x';\epsilon)&=&f_{eq}[g^{0a}_{\alpha\alpha}(x,x';\epsilon)-g^{0r}_{\alpha\alpha}(x,x';\epsilon)]  \\
     &+&                  [f_{S_\alpha}-f_{eq}][g^{0a}_{\alpha\alpha}(x,x';\epsilon)-g^{0r}_{\alpha\alpha}(x,x';\epsilon)],\no
\eea
where $f_{eq}(\epsilon)$ is the distribution function in  equilibrium and  thus the first term makes no contribution to current.

\subsection{ The retarded Green's functions of the upper edge with  a  side-attached quantum dot} \label{App:A(4)}

If a \textcolor{blue}{QD} with a localized level $\epsilon_d$ is side-attached to the upper edge, the Hamiltonian for the isolated upper edge $H_u$  is replaced by $H_u+H_d+H_{ud}$. The retarded Green's functions of the isolated upper edge  are denoted by $g^{0r}$ and  those coupled to a \textcolor{blue}{QD} by $g^r$. They are related  by ($ x'<x_d<x$)
\bea \label {eqsA30}
g^{r}_{uu}(x,x';\epsilon)=g^{0r}_{uu}(x,x';\epsilon)+g^{0r}_{uu}(x,x_d;\epsilon)t_d  g^{r}_{du}(x_d,x';\epsilon). \no \\
\eea
The mix Green's functions associated with the dot and the upper edge  $g^{r}_{du}(x_d,x';\epsilon)$ satisfy
\bea \label{eqsA31}
g^{r}_{du}(x_d,x';\epsilon)&=&g^{r}_{dd}(\epsilon)t_d g^{0r}_{uu}(x_d,x';\epsilon).
\eea

We also have equations for the retarded Green's function of the  \textcolor{blue}{QD}  $g^{r}_{dd}(\epsilon)$ and the mixed Green's function $g^{r}_{ud}(x_d,x_d;\epsilon)$
\bea
g^{r}_{dd}(\epsilon)&=&g^{0r}_{dd}(\epsilon)[1+t_d g^{r}_{ud}(x_d,x_d;\epsilon)],  \\\label{eqsA32}
g^{r}_{ud}(x_d,x_d;\epsilon)&=&g^{0r}_{uu}(x_d,x_d;\epsilon)t_d g^{r}_{dd}(\epsilon), \label{eqsA33}
\eea
where $g^{0r}_{dd}$ is the retarded Green's function of the isolated \textcolor{blue}{QD}. Now we solve for  $g^{r}_{dd}(\epsilon)$ by substituting  Eq.~\eqref{eqsA33} into \textcolor{blue} {(A32)} and find
\bea \label{eqsA34}
g^{r}_{dd}(\epsilon)=\frac{1}{[g^{0r}_{dd}(\epsilon)]^{-1}-\Sigma^{r}_{d}(\epsilon)}.
\eea

The retarded Green's function of the isolated \textcolor{blue} {QD} is 
$g^{0r}_{dd}(\epsilon)=(\epsilon-\epsilon_d+i0^+)^{-1}$, and the self-energy of the retarded Green's function of the \textcolor{blue} {QD} due to edge-dot coupling is given by
\bea \label{eqsA35}
\Sigma^{r}_{d}(\epsilon)=t_d^{2} g^{0r}_{uu}(x_d,x_d;\epsilon)=-i\frac{t_d^{2}}{2\hbar v_\text{F}}=-i\frac{\hbar v_\text{F}\gamma^2_d}{d}=-i\Gamma. \no \\
\eea 
Note that we have assumed $\theta(0)=1/2$ and thereby $g^r_{\alpha\alpha}(x,x;\epsilon)=\frac{1}{i2\hbar v_F}$. The retarded Green's function on the dot takes a simple expression
\begin{align} \label{eqsA36}
g^{r}_{dd}(\epsilon)=  \frac{1}{\epsilon-\epsilon_d+i\Gamma} .  
\end{align}

Plugging  Eqs.~\eqref{eqsA24},~\eqref{eqsA31}  and~\eqref{eqsA36} into Eq.~\eqref{eqsA30}, we get an explicit expression for the retarded Green's functions of the isolated upper edge in the presence of a side-coupled \textcolor{blue}{QD} for $x'<x_d<x$
\bea\label{eqsA37}
g^{r}_{uu}(x,x',\epsilon)=-\frac{i}{\hbar v_\text{F}}e^{i\epsilon(x-x')/\hbar v_\text{F}} \theta(x-x')\frac{\epsilon-\epsilon_d-i\Gamma}{\epsilon-\epsilon_d+i\Gamma}.\no\\   
\eea
The assumption $\theta(0)=1/2$ used to obtain the expression for the self-energy Eq.~\eqref{eqsA35} of the dot's Green's function is not trivial but even subtle.
If we view $\frac{\epsilon-\epsilon_d-i\Gamma}{\epsilon-\epsilon_d+i\Gamma}$ as the transmission amplitude $\tau_d(\epsilon)$ across the point $x_d$, the transmission probability must be always one, irrespective of the electron's incident energy, because there is no mechanism for the electron to reflect back. Otherwise we will find  a less-than-one transmission probability with an amplitude $\tau_d(\epsilon)=\frac{\epsilon-\epsilon_d}{\epsilon-\epsilon_d+i\Gamma}$  if assuming $\theta(0)=1$.  It contradicts the fact that chiral electrons always go through the point to which the \textcolor{blue}{QD} is side-attached.
In Appendix~\ref{App:C}, we will confirm  this result  using the scattering matrix approach.

The effect resulting from the presence of a side-attached \textcolor{blue}{QD} is incorporated into the Green's functions of the isolated upper edge as a transmission amplitude of unit modulus with an incident-energy-dependent phase, since the straight-through transmission probability across $x_d$ is obviously unity. Eq.~\eqref{eqsA37} may be repressed in the manner
\bea\label{eqsA38}
\hspace{-0.3cm}
\hspace{-0.4cm} g^{r}_{uu}(x,x',\epsilon)=-\frac{i}{\hbar v_\text{F}}e^{i\epsilon(x-x')/\hbar v_\text{F}} \theta(x-x')\tau_d(\epsilon), \label{eqs31}
\eea
for $x'<x_d<x$, and  $g^{r}_{uu}(x,x',\epsilon)=g^{r0}_{uu}(x,x',\epsilon)$ for other configurations,

\subsection{Explicit expressions for the full propagators}\label{App:A(5)}

We present explicit expressions for the various retarded Green's functions  in this section, which will be useful in the subsequent applications. For convenience, we denote the matrix functions appearing in Eqs.~\eqref{eqsA10},~\eqref{eqsA11},~\eqref{eqsA15} by $\hat{A},\hat{ B},\hat{ C}$, respectively: $\hat A=(\hat{1}-\hat{\Sigma}^{r}_{u}\hat g^r_{uu})^{-1}, \hat B=(\hat{1}-\hat g^r_{uu}\hat \Sigma ^{r}_{u})^{-1}, \hat C=(\hat{1}-\hat g^r_{ll}\hat \Sigma ^{r}_{l})^{-1}$. They are derived readily as
\begin{equation} \label{eqsA39}
\hat X= \begin{bmatrix}\frac{1}{1+\gamma_1^2 }& 0 \\
X_{21}(\epsilon) & \frac{1}{1+\gamma_2^2} \end{bmatrix},  X=A,B,C
\end{equation}
with non-diagonal elements given by
\bea
A_{21}(\epsilon)&=&\frac{-2\big[\gamma_2^2 \tau_d(\epsilon)e^{i\epsilon L_u/\hbar v_\text{F}}+\gamma_1\gamma_2e^{i(2\pi \frac{\Phi}{\Phi_0}+\epsilon  L_l/\hbar v_\text{F})}\big]}
 {(1+\gamma_1^2)(1+\gamma_2^2)} ,\no \\
 B_{21}(\epsilon)&=&\frac{-2\big[\gamma_1^2 \tau_d(\epsilon)e^{i\epsilon L_u/\hbar v_\text{F}}+\gamma_1\gamma_2e^{i(2\pi \frac{\Phi}{\Phi_0}+\epsilon  L_l/\hbar v_\text{F})}\big]}
 {(1+\gamma_1^2)(1+\gamma_2^2)} , \no \\
  C_{21}(\epsilon)&=&\frac{-2\big[\gamma_1^2e^{i\epsilon L_l/\hbar v_\text{F}} +\gamma_1\gamma_2\tau_d(\epsilon)e^{-i(2\pi \frac{\Phi}{\Phi_0}-\epsilon  L_u/\hbar v_\text{F})}\big]}
 {(1+\gamma_1^2)(1+\gamma_2^2)}, \no
\eea
where $\Phi$ is the magnetic flux enclosed by two arms of the \textcolor{blue}{MZFI} and $\Phi_0=hc/e$ is the  flux quantum. 

Substituting  Eqs.~\eqref{eqsA24},~\eqref{eqsA39}  into  Eq.~\eqref{eqsA11}, we obtain the following  two propagators $G^{r}_{uu}(x,x_i;\epsilon)$ ($x_1<x_2<x$) 
\bea  \label{eqsA40}
G^{r}_{uu}(x,x_2;\epsilon)= -\frac{i} {\hbar v_\text{F} (1+\gamma_2^2)}e^{i\epsilon(x-x_2)/\hbar v_\text{F}}, 
\eea
and
\bea\label{eqsA41}
G^{r}_{uu}(x,x_1;\epsilon)&=&-\frac{i\tau_d(\epsilon)}{\hbar v_\text{F}(1+\gamma_1^2) } e^{i\epsilon(x-x_1)/\hbar v_\text{F}} \no \\ &&-A_{21}(\epsilon)\frac{i}{\hbar v_\text{F}}e^{i\epsilon(x-x_2)/\hbar v_\text{F}}. 
\eea
It is these two retarded Green's functions that will be used to evaluate the current flowing into drain $D_1$ in Sec.~\ref{App:A(6)}.

Another four useful propagators are $G^r_{\alpha\beta}(x,x';\epsilon)$ $(x'<x_1<x_2<x)$, and are obtained after back substitution of Eqs. ~\eqref{eqsA24} and ~\eqref{eqsA39}  into  ~\eqref{eqsA12}-\eqref{eqsA15}, respectively
\begin{widetext}
\bea
G^r_{uu}(x,x';\epsilon)&=& \frac{e^{i\epsilon(x-x')/\hbar v_F}}{i\hbar v_F}\big[\sqrt{R_1R_2}\tau_d(\epsilon)-\sqrt{T_1T_2}e^{i(2\pi \frac{\Phi}{\Phi_0}+\epsilon \Delta L/\hbar v_\text{F})}\big], \\ \label{eqsA42}
G^r_{ll}(x,x';\epsilon)&=& \frac{e^{i\epsilon(x-x')/\hbar v_F}}{i\hbar v_F}\big[\sqrt{R_1R_2}-\sqrt{T_1T_2}\tau_d(\epsilon)e^{-i(2\pi \frac{\Phi}{\Phi_0}+\epsilon \Delta L/\hbar v_\text{F})}\big], \\ \label{eqsA43}
G^r_{lu}(x,x';\epsilon)&=& -\frac{e^{i\epsilon(x-x')/\hbar v_F}}{\hbar v_F}\big[\sqrt{R_1T_2}\tau_d(\epsilon)+\sqrt{T_1R_2}e^{i(2\pi \frac{\Phi}{\Phi_0}+\epsilon \Delta L/\hbar v_\text{F})}\big], \\ \label{eqsA44}
G^r_{ul}(x,x';\epsilon)&=& -\frac{e^{i\epsilon(x-x')/\hbar v_F}}{\hbar v_F}\big[\sqrt{R_1T_2}+\sqrt{T_1R_2}\tau_d(\epsilon)e^{-i(2\pi \frac{\Phi}{\Phi_0}+\epsilon \Delta L/\hbar v_\text{F})}\big], \label{eqsA45}
\eea
\end{widetext}
where $\delta L=L_l-L_u$ is the length difference of the lower and upper arms of the \textcolor{blue} {MZFI},
$T_i=4\gamma^2_i/(1+\gamma^2_i)^2$ and $R_i=(1-\gamma^2_i)^2/(1+\gamma^2_i)^2$ are the transmission  and reflection coefficients of carriers at \textcolor{blue} {QPC}i.

\subsection{Current flowing into drain $D_1$ }\label{App:A(6)}

In this section, we derive an explicit expression for the current $I_{D_1}$ flowing into $D_1$. One may choose the Fermi-Dirac distribution function $f_{D_1}=f(\epsilon-\mu_{D_1})=1/[e^{(\epsilon-\mu_{D_1})/k_BT}+1]$ of $D_1$ as the equilibrium distribution function $f_{eq}(\epsilon)$,  and obtains  after inserting the lesser Green's function~\eqref{eqsA20} of the  upper edge   into current expression~\eqref{current} 

\bea\label{eqsA46}
I_{D_1}&=&I^0_{D_1}+\frac e h  \int^{+\infty}_{-\infty}d\epsilon[f_{S_2}(\epsilon)-f_{S_1}(\epsilon)] i\hbar v_\text{F} \no \\
&& \sum_{i,j}G^{r}_{uu}(x,x_i;\epsilon) \Gamma^u_{ij}(\epsilon) G^{a}_{uu}(x_j,x;\epsilon),
\eea
where
\begin{align}\label{eqsA47}
I^0_{D_1}= \frac{e}{h}\int^{+\infty}_{-\infty}d\epsilon\big\{[f_{D_1}(\epsilon)-f_{S_1}(\epsilon)],  
\end{align}
is the direct current contributed by chiral electrons  injected from the  source $S_1$  without tunneling into the lower edge or the \textcolor{blue} {QD}.

 It is suggestive  to rewrite the above equation into the Landauer-B\"{u}ttiker form
\bea\label{eqsA48}
I_{D_1}&=&\frac e h \int^{+\infty}_{-\infty}d\epsilon \big\{ [f_{D_1}(\epsilon)-f_{S_1}(\epsilon)]\mathcal{T}_{D_1S_1}(\epsilon) +\no \\ &&[f_{D_1}(\epsilon)-f_{S_2}(\epsilon)]\mathcal{T}_{D_1S_2}(\epsilon)\big\}. 
\eea
In Eq.~\eqref{eqsA48},  $\mathcal{T}_{D_jS_i}(\epsilon), i,j=1,2$ are the transmission probabilities for chiral electrons tunneling through the device from  lead $S_i$ to lead $D_j$, and given by
\bea\label{eqsA49}
\mathcal{T}_{D_1S_i}(\epsilon)&=&\delta_{1i}+(-1)^{\delta_{1i}} \hbar v_\text{F} 
 \sum_{m,n}G^{r}_{uu}(x,x_m;\epsilon)\Gamma^u_{mn}(\epsilon)\no\\&& G^{a}_{uu}(x_n,x;\epsilon), \hspace{0.3cm} i=1,2.
\eea
where the coupling matrix elements are found to be 
\begin{align}\label{eqsA50}
\Gamma^u_{ij}(\epsilon)=4\hbar v_\text{F} \gamma_i\gamma_je^{-i(2\pi\frac{\Phi}{\Phi_0}+\epsilon L_l/\hbar v_\text{F})(1-\delta_{ij})}.
\end{align}

There are two contributions to the current flowing into $D_1$, obviously from in Eq. \eqref{eqsA48}: source $S_1$ and source $S_2$. From their definitions, we may infer that $G^r(x_i,x_j)=[G^{a}(x_j,x_i)]^*$ and $\Gamma_u(x_i,x_j)= [\Gamma_{u}(x_j,x_i)]^*$.  We calculate each term in Eq.~\eqref{eqsA49} and obtain
\begin{widetext}
\bea
\hbar v_\text{F}\Gamma^u_{22}(\epsilon)|G^r_{uu}(x,x_2;\epsilon)|^2&=&\frac{4\gamma^2_2}{(1+\gamma^2_2)^2}, \no  \\
\hbar v_\text{F}\Gamma^u_{11}(\epsilon)|G^r_{uu}(x,x_1;\epsilon)|^2&=&\frac{4\gamma^2_1}{(1+\gamma^2_1)^2(1+\gamma^2_2)^2}\Big[(1-\gamma^2_2)^2|\tau_d(\epsilon)|^2+
4\gamma^2_1\gamma^2_2-4\gamma_1\gamma_2(1-\gamma^2_2) \no \\
 && Re\big[\tau_d(\epsilon)e^{-i(2\pi\frac{\Phi}{\Phi_0}+\epsilon\Delta L/\hbar v_\text{F})}\big]\Big], \no \\
\hbar v_\text{F} Re\big[\Gamma^u_{12} (G^r_{uu}(x,x_1;\epsilon)G^a_{uu}(x_2,x;\epsilon)\big])&=&\frac{4\gamma_1\gamma_2}{(1+\gamma^2_1)(1+\gamma^2_2)^2}\Big[-2\gamma_1\gamma_2+(1-\gamma^2_2)Re
\big[\tau_d(\epsilon)e^{-i(2\pi\frac{\Phi}{\Phi_0}+\epsilon\Delta L/\hbar v_\text{F})}\big]\Big]. \no 
\eea
\end{widetext}

Substituting the above three terms back into Eq.~\eqref{eqsA49} and regrouping similar terms, the transmission probability is found to be  the squared magnitude of a transmission amplitude 
\bea\label{eqsA51}
\mathcal{T}_{D_1S_2}(\epsilon)&=&|\sqrt{R_1T_2}+\sqrt{T_1R_2}\tau_d(\epsilon) e^{-i(\phi_{AB}+\epsilon \Delta L/\hbar v_\text{F})}|^2, \no \\
\mathcal{T}_{D_1S_1}(\epsilon)&=&1-\mathcal T_{D_1S_2}(\epsilon). 
\eea

\section{Current expression in the presence of contact hamiltonian}\label{App:B}

In this section, we consider a standard set up: a central scattering region(the device) is connected via  leads to contacts as reservoirs. The Hamiltonian can be decomposed as $H=H_C+H_{D}+H_T$, where $H_{D}$  is the device Hamiltonian given in the main text, $H_C$ represents the Hamiltonian of the contacts, and $H_T$ is the coupling Hamiltonian between the device and the leads. The leads are modeled by one-dimensional chiral edge states and  described by the chiral Hamiltonian
\begin{align}\label{eqsB1}
H_C=-i\sum_{\alpha=u,l}^{i=S,D}\!\hbar v_F^{i_\alpha} \int dy \hat{\varphi}^\dagger _{i_\alpha}(y) \partial_x \hat{\varphi}_{i_\alpha}(y), 
\end{align}
 where $\hat{\varphi}_{i_\alpha}(y)$ are the field operators for  lead $i_\alpha$. The chemical potential of contact $i_\alpha$ is $\mu_{i_\alpha}$, where $i=S,D$ and $\alpha=u,l$ (or $ 1,2 $).  The couplings between the leads and the device are given by the tunneling Hamiltonian 
 \begin{align} \label{eqsB2}
  H_T=\sum_{\alpha=u,l}^{i=S,D}[t_{i\alpha}\hat{\varphi}^{\dagger}_{i\alpha}(y_L)\hat{\psi}_{\alpha}(x_L)+H.c.], 
\end{align}
where $x_L$, $x_R$  and $y_L$, $y_R$ are, respectively, the positions at which the central device and the leads are connected.

The current flowing into the upper drain $D_1$  can be calculated from the time evolution of the number operator of the upper drain $D_1$:
\bea\label{eqsB3}
I_{D_1}&=&e\frac{d}{dt}\int dy \big<\hat{\varphi}^{\dagger}_{D_u}(y,t)\hat{\varphi}_{D_u}(y,t)\big>_H \no\\
&=&-\frac{ie}{\hbar}
\big[t_{Du}\big<\hat{\varphi}^{\dagger}_{D_u}(y_R,t)\hat{\psi}_u(x_R,t)\big>_H-H.c.\big].
\eea
With the definition of the lesser Green's function $\mathcal{G}^<_{uD}(x,t';y,t)=\frac{i}{\hbar}\big<\hat{\varphi}^{\dagger}_{D_u}(y,t)\hat{\psi}_u(x,t')\big>_H$, the above equation can be rewritten as
\begin{align}\label{eqsB4}
I_{D_1}=-2e Re[t_{Du}\mathcal{G}^<_{uD}(x_R,t;y_R,t)].
\end{align} 

The equation of motion for the mixed retarded Green's function $\mathcal{G}^r_{uD}(x_R,y_R;\epsilon)$ gives
\bea\label{eqsB5}
\hspace{-0.5cm}
 \mathcal{G} ^{r}_{uD}(x_R,y_R;\epsilon)= \mathcal{ G}^r_{uu}(x_R,x_R;\epsilon) t_{Du}^{*}g^{r}_{D_u}(y_R, y_R,\epsilon).
\eea

The corresponding lesser Green's function is found by applying the Langreth rules 
\bea \label{eqsB6}
 \mathcal{G}^{<}_{uD}(x_R,y_R;\epsilon)&=&\mathcal{ G}^r_{uu}(x_R,x_R;\epsilon) t_{Du}^{*}g^{<}_{D_u}(y_R, y_R,\epsilon)+\no\\
 && \mathcal{G}^<_{uu}(x_R,x_R;\epsilon) t_{Du}^{*}g^{a}_{D_u}(y_R, y_R,\epsilon).
\eea

Eq.~\eqref{eqsB3} becomes after inserting Eq.~\eqref{eqsB6} into Eq.~\eqref{eqsB3} 
\bea\label{eqsB7}
I_{D_1}&=&-2e\int \frac{d\epsilon}{2\pi\hbar}  Re\big[\mathcal{G}^r_{uu}(x_R,x_R;\epsilon)\Xi^{<}_{u}(y_R, y_R,\epsilon)+ \no\\
&&\mathcal{G}^<_{uu}(x_R,x_R;\epsilon)\Xi^{a}_{u}(y_R,y_R;\epsilon)\big],
\eea 
where $\Xi^{r,a,<}_{u}(x_m,x_m;\epsilon)=|t_{mu}|^2g^{r,a,<}_{mu}(y_m,y_m;\epsilon), m=L,R$ are various self-energies arising from the lead-upper-edge couplings. With the relation $G^r-G^a=G^>-G^<$, the above equation may be recast in another form 
\bea\label{eqsB8}
I_{D_1}&=&\frac{e}{h}\int d\epsilon  \big[\Xi^{<}_{u}(y_R, y_R,\epsilon)\mathcal{G}^>_{uu}(x_R,x_R;\epsilon)-\no\\
&&\Xi^{>}_{u}(y_R,y_R;\epsilon)\mathcal{G}^<_{uu}(x_R,x_R;\epsilon) \big].
\eea 

In the presence of contacts, the Dyson's equation for the  retarded Green's function $\mathcal{G}^r_{uu}(x,x';\epsilon)$ is
\bea\label{eqsB9}
 \mathcal{G}^{r}_{uu}(x,x';\epsilon)&=&\mathcal{G}^{0r}_{uu}(x,x';\epsilon)+ \sum_{m=L,R} \mathcal{G}^{0r}_{uu}(x, x_m,\epsilon) \no \\&& \Xi^{r}_{u}(x_m,x_m;\epsilon)  \mathcal{G}^r_{uu}(x_m,x';\epsilon), 
 \eea
 where $\mathcal{G}^{0}_{uu}$ denotes the Green's function of the upper-edge in the absence of the lead-upper-edge couplings. 
From this equation one can obtain the following Keldysh equation
\begin{widetext}
\bea\label{eqsB10}
\mathcal{G}^{</>}_{uu}(x,x')&=&\sum_{m,n=L,R}\big[\delta_{x,x_m}+\mathcal{G}^r_{uu}(x,x_m)\Xi^r_{u}(x_m,x_m)\big]
                  \mathcal{G}^{0</>}_{uu}(x_m,x_n)\big[\delta_{x,x_m}+\Xi^a_{u}(x_n,x_n)G^a_{uu}(x_n,x)\big]  \no  \\
                     & &\hspace{2cm} +\sum_{m=L,R} \mathcal{G}^r_{uu}(x,x_m) \Xi^{</>}_{u}(x_m,x_m)\mathcal{G}^a_{uu}(x_m,x).
\eea
\end{widetext}
The Green's function $\mathcal{G}^{0}_{uu}$ reduces to the Green's function $G_{uu}$ if the lower edge is decoupled from the source $S_l$($S_2$) and the drain $D_l$($D_2$). One may get the Green's function $\mathcal{G}^{0}_{uu}$  from $G_{uu}$ after replacing the decoupled Green's functions $g^{0}_{ll}$ from the lower contacts $S_l,D_l$ by the  coupled Green's functions $g_{ll}$ to the lower contacts. Equation of motion for  $g^r_{ll}$ yields
 \bea \label{eqsB11}
 g^r_{ll}(x,x';\epsilon)&=&g^{0r}_{ll}(x,x';\epsilon)+\!\sum_{m=L,R}g^{0r}_{ll}(x,x_m;\epsilon)\Xi_{l}^r(y_m,y_m)\no\\&&
 g^{r}_{ll}(x_m,x';\epsilon), 
 \eea
where $\Xi^{r}_{l}(x_m,x_m;\epsilon)=|t_{ml}|^2g^r_{ml}(y_m,y_m;\epsilon), m=L,R$ are the retarded self-energies due to the lead-lower-edge couplings.  It follows from Eq.~\eqref{eqsB11} that $g^r_{ll}(x_i,x_j;\epsilon)=g^{0r}_{ll}(x_i,x_j;\epsilon), i,j=1,2$, $x_L<x_1,x_2<x_R$,
and $\mathcal{G}^{0}_{uu}(x,x')=G^r_{uu}(x,x')$.  It justifies that the formalism in Appendix~\ref{App:A} to solve for full propagators through the device is reasonable from  the propagators of the isolated upper and lower edges from contacts, even if we made an assumption that the edges are connected to the contacts.  The corresponding Keldysh equation to Eq.~\eqref{eqsB11} reads

\bea \label{eqsB12}
\hspace{-0.3cm}
 g^<_{ll}(x,x')=\sum_{m=L,R}g^{r}_{ll}(x,x_m)\Xi_{l}^<(y_m,y_m)g^{a}_{ll}(x_m,x'),\no \\
 \eea
 where $\Xi^{r,a,<}_{l}(x_m,x_m;\epsilon)=|t_{ml}|^2g^{r,a,<}_{ml}(y_m,y_m;\epsilon), m=L,R$. 
 In addition, we have 
 \bea\label{eqsB13}
 \hspace{-0.3cm}
 \mathcal{G}^{0<}_{uu}(x,x')=\sum_{i,j}\mathcal{G}^{0r}_{uu}(x,x_i)\Sigma^{<}_{u,ij} (\epsilon)G^{0a}_{uu}(x_j,x'),
 \eea 
where $\Sigma^{<}_{u,ij}(\epsilon)=t_it^*_jg^<_{ll}(x_i,x_j;\epsilon)$ is the lesser self-energy arising from the tunneling from the upper edge to the lower edge which is coupled to the contacts.  Substituting Eq.~\eqref{eqsB12} and ~\eqref{eqsB13} into Eq.~\eqref{eqsB10}, 
Eq.~\eqref{eqsB10} becomes 
\bea\label{eqsB14}
\mathcal{G}^{</>}_{uu}(x,x')&=&\sum_{m,n}\mathcal{ G}^r_{ul}(x,x_m)\Xi^{</>}_{l}(x_m,x_m) \mathcal{G}^a_{lu}(x_m,x')+\no\\&&\sum_{m=L,R} \mathcal{G}^r_{uu}(x,x_m) \Xi^{</>}_{u}(x_m,x_m)\mathcal{G}^a_{uu}(x_m,x).\no\\
\eea
 Here we have used the relations: 
 \bea\label{eqsB15}
 \begin{split}
 \mathcal{G}^r_{ul}(x,x')=\sum_i \mathcal{G}^r_{uu}(x,x_i)t_i g^r_{ll}(x_i,x'),\\
 \mathcal{G}^a_{lu}(x,x')=\sum_{j}g^a_{ll}(x,x_j)t^*_j \mathcal{G}^a_{uu}(x_j,x'). 
 \end{split}
 \eea  
Similar to Eqs.~\eqref{eqsA24}, ~\eqref{eqsA25} and ~\eqref{eqsA29} in Appendix \ref{App:A(3)}, the various  Green's functions of the lead $m\alpha$ ($m=L,R 
$ or $ S,D$; $\alpha=u,l$ or $1,2 $) are given by 
\begin{equation} \label{eqsB16}
\begin{split}
g^{a/r}_{m\alpha}(y_m,y_m;\epsilon)&=\frac{\pm i}{2\hbar v_\text{F}^{m\alpha}}, \\
 g^{<}_{m\alpha}(y_m,y_m;\epsilon)&=\frac{i}{\hbar v_\text{F}^{m\alpha}}f_{m\alpha}(\epsilon), \\ g^{>}_{m\alpha}(y_m,y_m;\epsilon)&=\frac{i}{\hbar v_\text{F}^{m\alpha}}[1-f_{m\alpha}(\epsilon)]. 
 \end{split}
 \end{equation}
If we set the edge-lead coupling constant as $t_{m\alpha}=2\hbar \sqrt{v_Fv_F^{m\alpha}}\gamma_{m\alpha}, m=L,R, \alpha=u,l$, then the various self-energies due to edge-lead couplings are 
\begin{equation} \label{eqsB17}
\begin{split}
\Xi^{a/r}_{\alpha} (y_m,y_m)&=\pm i2\hbar v_F \gamma^2_{m\alpha},\\
 \Xi^{<}_{\alpha} (y_m,y_m)&=i4\hbar v_F \gamma^2_{m\alpha}f_{m\alpha}(\epsilon), \\
  \Xi^{>}_{\alpha} (y_m,y_m)&=i4\hbar v_F \gamma^2_{m\alpha}[1-f_{m\alpha}(\epsilon)]. 
  \end{split}
\end{equation}
Substituting Eq.~\eqref{eqsB14} and ~\eqref{eqsB17} into Eq.~\eqref{eqsB8}, we find
\begin{widetext}
\bea\label{eqsB18}
\hspace{-0.6cm}
I_{D_1}=\frac{e}{h} (4\hbar v_F)^2 \int d\epsilon  \big\{[f_{D_1}(\epsilon)-f_{S_1}(\epsilon)](\gamma_{S_1} \gamma_{D_1})^2
|\mathcal{G}^r_{uu}(x_R,x_L)|^2+ [f_{D_1}(\epsilon)-f_{S_2}(\epsilon)](\gamma_{S_2} \gamma_{D_1})^2|\mathcal{G}^r_{ul}(x_R,x_L)|^2\big\}.
\eea
\end{widetext}
To obtain Eq.~\eqref{eqsB18} we have used the fact that $\mathcal{G}^{r,a}_{ul}(x_m,x_
m)=0,m=L,R$ for chiral electrons. Following the same procedure as we did in Appendex \ref{App:A(1)}, it is useful to rewrite the Dyson's equation Eq.~\eqref{eqsB9} for the retarded Green's functions $\mathcal{G}^r_{uu}(x_m,x_n), m,n=L,R$ in matrix form $\hat{\mathcal{G}}^r_{uu}= \hat{\mathcal{G}}^{0r}_{uu}+ \hat{\mathcal{G}}^{0r}_{uu} \hat{\Xi}^r_u \hat{\mathcal{G}}^r_{uu}$ with a diagonal self-energy matrix $\hat{\Xi}^r_u={\it dia}[\hat{\Xi}^r_u(x_L,x_L),\hat{\Xi}^r_u(x_R,x_R)]$.  Solving this matrix equation gives
\bea\label{eqsB19}
\hspace{-0.5cm}
\mathcal{G}^r_{uu}(x_R,x_L)&=&\frac{1}{(1+\gamma_{S_1}^2)(1+\gamma_{D_1}^2)}\mathcal{G}^{0r}_{uu}(x_R,x_L)\no\\
&=&\frac{1}{(1+\gamma_{S_1}^2)(1+\gamma_{D_1}^2)}G^{r}_{uu}(x_R,x_L). 
\eea
Also we have 
\bea \label{eqsB20}
\mathcal{G}^r_{uu}(x_R,x')&=&\sum_{n=L,R}\big[(\hat{1}-\hat{\mathcal{G}}^{0r}_{uu}\hat{\Xi}^r_u)^{-1}\big]_{Rn} \mathcal{G}^{0r}_{uu}(x_n,x')\no\\
&=&\sum_{n=L,R}\big[(\hat{1}-\hat{\mathcal{G}}^{0r}_{uu}\hat{\Xi}^r_u)^{-1}\big]_{Rn} G^{r}_{uu}(x_n,x'). \no\\
\eea
To solve for $ \mathcal{G}^r_{ul}(x,x_L)$, we also need  to know $g^r_{ll}(x,x_L)$, which is found from Eq.~\eqref{eqsB11} to be 
\bea \label{eqsB21}
\hspace{-0.5cm}
g^r_{ll}(x,x_L)=\sum_{n=L,R} g^{0r}_{ll}(x,x_n)\big[(\hat{1}-\hat{\Xi}^r_l\hat{g}^{0r}_{ll})^{-1}\big]_{nL}. 
\eea
We finally obtain after plugging Eqs.~\eqref{eqsB20} and ~\eqref{eqsB21} into Eq.~\eqref{eqsB15} 
\begin{widetext}
\bea \label{eqsB22}
\hspace{-0.4cm}
\mathcal{G}^r_{ul}(x_R,x_L)=\sum_{m,n=L,R}\big[(\hat{1}-\hat{\mathcal{G}}^{0r}_{uu}\hat{\Xi}^r_u)^{-1}\big]_{Rm}G^r_{ul}(x_m,x_n)
 [(\hat{1}-\hat{\Xi}^r_l\hat{g}^{0r}_{ll})^{-1}\big]_{nL}=\frac{1}{(1+\gamma_{S_2}^2)(1+\gamma_{D_1}^2)}G^{r}_{ul}(x_R,x_L). 
\eea
\end{widetext}
With Eqs.~\eqref{eqsB20} and ~\eqref{eqsB22} in hand, the current in Eq.~\eqref{eqsB18} reduces to 
\bea\label{eqsB23}
\begin{split}
I_{D_1}&=\frac{e}{h}  \int d\epsilon \Big\{[f_{D_1}(\epsilon)-f_{S_1}(\epsilon)]T_{S_1}\mathcal{T}_{D_1S_1}(\epsilon)T_{D_1}\\
&+[f_{D_1}(\epsilon)-f_{S_2}(\epsilon)]T_{S_2}\mathcal{T}_{D_1S_2}(\epsilon)T_{D_1}\Big\},
\end{split}
\eea
where   $T_{m\alpha}= 4\gamma^2_{m\alpha}/(1+\gamma^2_{m\alpha})^2$ is the transmission coefficient through the lead-edge contact $m\alpha$, and $\mathcal{T}_{D_1S_i}(\epsilon)=|s_{D_1S_i}|^2=|i\hbar v_F G^r_{D_1S_i}|^2=|i\hbar v_F G^r_{u\alpha}(x_R,x_l)|^2,i=1,2,\alpha=u,l$ is the transmission coefficient from the top or bottom left to the top right of the device and can be easily understood from the Fisher-Lee relation.

If the edges are connected to the leads ballistically, i.e.,
 the outer segments of edges are considered as leads, $\gamma_{m\alpha}=1$, and thus $T_{S_i}=T_{D_1}=1$, the current expression 
 Eq.~\eqref{eqsB23} is the same as Eq.~\eqref{eqsA48} in Appendix~\ref{App:A(6)}. 
 
 From Eq.~\eqref{eqsB23}, one sees that each distribution function can be decomposed into  $f_{m\alpha}=[f_{m\alpha}-f_{eq}]+f_{eq}$, and $f_{eq}$ does not contribute to the current, since the current depends only on the difference of distribution functions. In addition, due to chiral nature of electron motion, the current flowing into a drain contact does not depend on the other drain contact. Therefore it is reasonable to omit the contribution from the drains when considering the lesser Green's functions of the edges, as we did in Appendix~\ref{App:A}.

\section{Transmission amplitudes through $x_d$: A scattering matrix approach }\label{App:C}

In this section, we derive the transmission amplitude for chiral electrons passing through $x_d$ on the upper edge based on the scattering matrix approach and  justify the assumption $\theta(0)=1/2$ in Appendix~\ref{App:A(3)}.

If the \textcolor{blue} {QD} is absent,  the transmission amplitude is obviously one. When the \textcolor{blue} {QD} is side-attached to the upper edge,  the Hamiltonian becomes $H=H_u+H_d+H_{ud}$. The Heisenberg equations of motion $i\hbar\partial_t \hat A=[\hat A,H]$ for $\hat{\psi}_u(x,t)$ and $\hat d(t)$ yield the following set of differential equations
\bea
i\hbar\partial_t \hat{\psi}_u(x,t)&=&-i\hbar v_F \partial_x \hat{\psi}_u(x,t)+t_d\hat{d}(t)\delta(x-x_d), \no \\
i\hbar \partial_t \hat{d}(t) &=& \epsilon_d\hat{d}(t)+t_d\hat{\psi}_u(x_d,t).\no
\eea
Fourier transforming the above differential equations, we obtain
\begin{align}\label{eqsC1} 
\epsilon \hat{\psi}_u(x,\epsilon)=-i\hbar v_F \partial_x \hat{\psi}_u(x,\epsilon)+t_d\hat{d}(\epsilon)\delta(x-x_d),  
\end{align}
\begin{align}\label{eqsC2}
\epsilon \hat{d}(\epsilon) = \epsilon_d \hat{d}(\epsilon)+t_d\hat{\psi}_u(x_d,\epsilon).
\end{align}
Integrating both sides of Eq.~\eqref{eqsC1}  from $x^-_d$ to $x^+_d$, we obtain
\begin{align} \label{eqsC3}
-i\hbar v_F\big [\hat{\psi}_u(x^+_d,\epsilon)-\hat{\psi}_u(x^-_d,\epsilon)\big]+t_d\hat{d}(\epsilon)=0. 
\end{align}
Using the fact $\hat{\psi}_u(x_d,\epsilon)=\big [\hat{\psi}_u(x^+_d,\epsilon)+\hat{\psi}_u(x^-_d,\epsilon)\big]/2$, and solving Eqs.~\eqref{eqsC2} and~\eqref{eqsC3} simultaneously,  one arrives at 
\begin{align}\label{eqsC4}
\hat{\psi}_u(x^+_d,\epsilon)=\frac{\epsilon-\epsilon_d-i\Gamma}{\epsilon-\epsilon_d+i\Gamma}\hat{\psi}_u(x^-_d,\epsilon), 
\end{align}
from which, one extracts the transmission amplitude  $\tau_d(\epsilon)=\frac{\epsilon-\epsilon_d-i\Gamma}{\epsilon-\epsilon_d+i\Gamma}$, same as the result in Appendix~\ref{App:A(4)}, thus justifying that
the assumption $\theta(0)=1/2$ is reasonable.

\section{Scattering matrix of the Mach-Zehnder-Fano interferometer}\label{App:D}

The Fisher-Lee relation\cite{Fisher,Datta} expresses the scattering matrix's element in terms of the retarded Green's function,  demonstrating the equivalence between the Landauer-B\"{u}ttiker scattering matrix approach and the Green's function formalism. The propagator $G^r_{pq}$ from $q$ to $p$ is related to the transmission amplitude $s_{qp}$ from lead $q$ to lead $p$ by the formula $s_{qp}=i\hbar\sqrt{v_pv_q}G^r_{pq}$, where $v_{p/q}$ represents the velocity of carriers in lead $p/q$. The scattering matrix in the current four-terminal setup is a $4\times4$ matrix with the structure 
\bea\label{eqsD1}
\hat s=\begin{pmatrix} 0 & 0 & s_{S_1D_1} & 0 \\ 0 & 0 &0 & s_{S_2D_2}\\ s_{D_1S_1} & s_{D_1S_2} & 0 & 0 \\ s_{D_2S_1} & s_{D_2S_2} & 0 &0 \end{pmatrix}, 
\eea 
where
\begin{widetext}
\bea
s_{D_1S_1}&=& \big[\sqrt{R_1R_2}\tau_d(\epsilon)-\sqrt{T_1T_2}e^{i(2\pi \frac{\Phi}{\Phi_0}+\epsilon \Delta L/\hbar v_\text{F})}\big]e^{i\epsilon(x-x')/\hbar v_F}, \\ \label{eqsD2}
s_{D_2S_2}&=&\big[\sqrt{R_1R_2}-\sqrt{T_1T_2}\tau_d(\epsilon)e^{-i(2\pi \frac{\Phi}{\Phi_0}+\epsilon \Delta L/\hbar v_\text{F})}\big]e^{i\epsilon(x-x')/\hbar v_F}, \\ \label{eqsD3}
s_{D_2S_1}&=& -i\big[\sqrt{R_1T_2}\tau_d(\epsilon)+\sqrt{T_1R_2}e^{i(2\pi \frac{\Phi}{\Phi_0}+\epsilon \Delta L/\hbar v_\text{F})}\big]e^{i\epsilon(x-x')/\hbar v_F}, \\ \label{eqsD4}
s_{D_1S_2}&=& -i\big[\sqrt{R_1T_2}+\sqrt{T_1R_2}\tau_d(\epsilon)e^{-i(2\pi \frac{\Phi}{\Phi_0}+\epsilon \Delta L/\hbar v_\text{F})}\big]e^{i\epsilon(x-x')/\hbar v_F}, \label{eqsD5} 
\eea
\end{widetext}
form a $2\times 2$ unitary scattering matrix, satisfying $ |s_{D_1S_1}|^2=|s_{D_2S_2}|^2$, $|s_{D_1S_2}|^2=|s_{D_2S_1}|^2$ , $\sum_{p'} s^*_{pq'}s_{qq'}=\delta_{pq}$; and 
\begin{align} \label{eqsD6}
s_{S_1D_1}=e^{-i\epsilon L_1/\hbar v_F}, \hspace{0.1cm} s_{S_2D_2}=e^{-i\epsilon L_2/\hbar v_F},
\end{align}
with  $L_1$ and $L_2$ being the lengths of two outer paths connecting the source and drain contacts. It is straightforward to check that the scattering matrix Eq.~\eqref{eqsD1} is unitary. Following the scattering matrix theory of shot noise in multi-terminal mesoscopic conductors\cite{Blanter,Buttiker2}, the components of the shot noise power tensor are 
\bea\label{eqsD7}
S_{pq}&=&\frac{e^2}{h}\int d\epsilon \sum_{p'\neq q'}s^*_{pq'}s_{pp'}s^*_{qp'}s_{qq'}[f_{q'}(1-f_{p'})+f_{p'}\no\\
&&(1-f_{q'})],\hspace{0.3cm} p,p',q,q'=S_1,S_2,D_1,D_2, 
\eea
where $f_p=\theta(\mu_p-\epsilon)$ is the zero temperature limit of the Fermi-Dirac distribution function in contact $p$. One can readily check that $S_{S_1S_1}=S_{S_2S_2}=0$, which is consistent with the fact that the electron stream injected from either source is not fluctuating at zero temperature, and 
\bea 
S_{D_1D_1}&=&\frac{2e^2}{h}\int^{eV}_0 \mathcal T_{D_1S_1}(\epsilon) \mathcal T_{D_1S_2}(\epsilon), \\ \label{eqsD8}
S_{D_2D_2}&=&\frac{2e^2}{h}\int^{eV}_0 \mathcal T_{D_2S_1}(\epsilon) \mathcal T_{D_2S_2}(\epsilon).  \label{eqsD9}
\eea
The transmission probabilities $\mathcal T_{pq}=|s_{pq}|^2$. Using the unitary property of the scattering matrix $\sum_{r} s^*_{pr}s_{rq}=\delta_{pq}$,  the nondiagonal components, $S_{pq}, p\neq q$, can be rewritten as 
\begin{align}\label{eqsD10}
S_{pq}=-\frac{2e^2}{h}\int d\epsilon \sum_{p'}s^\dagger_{pp'}s_{qp'}f_{p'} \sum_{q'} s^\dagger_{qq'}s_{pq'}f_{q'}, 
\end{align}
which is always negative due to the anti-bunching effect arising from Pauli exclusion principle\cite{Buttiker1,Buttiker2,Martin}. Using $f^2_p=f_p$, we obtain
\begin{align}\label{eqsD11}
S_{D_1D_2}=S_{D_2D_1}=-\frac{2e^2}{h}\int^{eV}_0 \mathcal T_{D_1S_1}(\epsilon) \mathcal T_{D_1S_2}(\epsilon). 
\end{align}
These components are related to one another by $S_{D_1D_1}=S_{D_2D_2}=-S_{D_1D_2}=-S_{D_2D_1}$.

%%%%%%%%%%%%%%%%%%%%%%%%%%%%%%%%%%%%%%%%%%%%%%%%%%%%%%%%

\end{document}